\pdfoutput=1
\documentclass[10pt,aps,prd,showpacs,twocolumn,superscriptaddress,nofootinbib,amsmath,amssymb]{revtex4-2}
 \usepackage{amsmath}
 \usepackage{amssymb}
 \usepackage{multirow}
 \usepackage{caption}
 \usepackage{subcaption}
 \usepackage{float}
\usepackage{graphicx}
\usepackage{txfonts}
\usepackage{soul}
\usepackage{url}
\usepackage[utf8]{inputenc}

\begin{document}


\title{Equilibrium models of Weyssenhoff spin fluid accretion tori around Kerr black holes}

\author{Sergio Gimeno-Soler}
\affiliation{Departamento de Matem\'atica da Universidade de Aveiro and
Center for Research and Development in Mathematics and Applications (CIDMA),
Campus de Santiago, 3810-183 Aveiro, Portugal.}

\author{Sayantani Lahiri}
\affiliation{Institut für Theoretische Physik und Astrophysik, Universität Würzburg, Emil-Fischer-Strasse 31, 97074 Würzburg, Germany}
\affiliation{University of Bremen, Center of Applied Space Technology and Microgravity (ZARM), 28359 Bremen.}

\author{Claus Lämmerzahl}%
\affiliation{University of Bremen, Center of Applied Space Technology and Microgravity (ZARM), 28359 Bremen.}



\begin{abstract}
The construction of equilibrium models of accretion disks around compact objects has become a highly relevant topic in the recent times, thanks to the current understanding that indicates a direct relationship between these objects with the electromagnetic emission of supermassive compact objects residing at center of the galaxies M87 and Milky Way both of which were recently observed by the Event Horizon Telescope Collaboration. As the physical properties of the compact sources are estimated using the results of computer simulations of the system comprising of the disk plus the compact object, adding new physical ingredients to the initial data of the simulation is pertinent to enhance our knowledge about these objects.
In this work, we thus present equilibrium solutions of geometrically thick, non-self-gravitating, constant orbital specific angular momentum, neutral Weyssenhoff spin fluid accretion tori in the Kerr spacetime, building upon a previous work 
that was restricted to the Schwarzschild geometry. Our models are obtained under the usual assumptions of stationarity and axisymmetry in the fluid quantities, circularity of the flow and a polytropic equation of state. We study how the deviations from an ideal no-spin fluid depend on both the magnitude of the macroscopic spin of the fluid and on the spin parameter of the Kerr black hole, carefully encompassing both the co-rotating and the counter-rotating cases. Our results demonstrate that the characterstic radii, the thickness and the radial extent of such a torus can change importantly in the presence of the macroscopic spin of the ideal fluid. We also find some limitations of our approach that constraint the amount of spin the fluid can have in the rotating Kerr background. Finally, we present a parameter space exploration that gives us additional constraints on the possible values of the fluid spin denoted by the parameter $s_0$.
\end{abstract}

\maketitle

\section{Introduction}

The existence of supermassive black holes (SMBHs) at the galactic centers is nearly unequivocally established by precision experiments ~\cite{GRAVITY:2018ofz} and~\cite{Ghez:1998ph} and a series of works~\cite{Magorrian:1997hw, Volonteri:2010wz, Kormendy:2013dxa}.
On the one hand, the growth of such black holes (BHs) is thought to be strongly favoured by accretion of surrounding plasma matter associated with emission of relativistic jets and outflows, on the other hand, accretion processes are also related to highly energetic gamma-ray bursts and X-ray binaries (for instance~\cite{1972Natur.235...37W, 1986ApJ...308..110M} correspond to earliest observations of stellar-mass BHs). 
Recently, first-time images of the close environment of a SMBH at the center of M87 galaxy~\cite{EventHorizonTelescope:2019dse, EventHorizonTelescope:2019pgp, EventHorizonTelescope:2019ggy,EventHorizonTelescope:2019pgp} and that of Sagittarius A$^{*}$~\cite{EventHorizonTelescope:2022xnr, EventHorizonTelescope:2022urf, EventHorizonTelescope:2022xqj,EventHorizonTelescope:2024hpu} at the center of Milky Way were published by the Event Horizon Telescope Collaboration together with the inferred physical parameters of both the central object and the surrounding astrophysical plasma. In order to obtain an estimation of the physical parameters of the source, the EHT Collaboration relies on a large number of general relativistic magnetohydrodynamics simulations of accretion disks around different compact objects and general relativistic ray-tracing codes to obtain the observational appearance of the simulated physical system. Therefore, increasing the number of available initial data through computation of new, well-motivated equilibrium solutions  thereby potentially improves the reliability of the physical interpretation of the observational data. \\
\indent Out of several accretion disk models proposed in the literature~\cite{AbramowiczFragile2013}, in the present paper, we have adopted the model of geometrically thick disks (tori), also known as Polish doughnuts~\cite{AbramowiczJaroszynskiSikora78} having a toroidal topology.
For stationary and axisymmetric spacetimes, the rotating non-self gravitating tori modelled by an ideal hydrodynamical fluid gives rise to exact analytical solutions of the relativistic Euler equation.
Notably, the shape and size of these disks are purely governed by the specific orbital angular momentum distribution and gradients of the fluid pressure. \\
\indent Nevertheless, the constituent fluid elements of such a torus may also possess intrinsic spin angular momentum attributed to the quantum nature of microscopic particles of the fluid elements, in that case, it culminates into a new form of the energy momentum tensor after incorporating the non-zero spin of the fluid.
In the continuum limit, the corresponding hydrodynamical theory is then obtained using an averaging approach from the microscopic theory of matter with non-zero spin \cite{1960PThPh..24..291H}. 
In this regard, the ideal Weyssenhoff spin fluid represents a phenomenological model providing a hydrodynamic description of matter involving an intrinsic spin degree of freedom~\cite{Weyssenhoff:1947iua} and the corresponding fluid elements are characterized by an intrinsic spin angular momentum proportional to their volume.
In relativity, it is known that the reference point of the spinning object, which usually taken as the center of mass (CM) depends on the observer. Since the intrinsic spin angular momentum of a such an object relies on the location of its CM about which it is evaluated, the spin, as a result, also becomes an observer-dependent quantity. Both the CM and the intrinsic spin are thus described by a rank-two antisymmetric spin tensor $S^{\mu \nu}$. While the purely spatial components (flux dipole) coincides with the three-density of spin of the matter in the rest frame, other spacetime components are assumed to be zero in the rest frame which in the covariant form leads to the specification of a spin supplementary condition (SSC) which in general is not unique and several choices of SSCs exist in the literature \cite{1956AcPP...15..389P,Frenkel:1926zz,Tulczyjew,Dixon:1970zza,newton1949localized,Kyrian:2007zz}.   \\
\indent The Weyssenhoff spin fluid model is formulated with Frankel SSC~\cite{Weyssenhoff:1947iua}. The corresponding Lagrangian theory of the ideal Weyssenhoff spin fluid has been put together by several authors \cite{Ray:1982qq, deOliveira:1991ja,Minkevich-Karakura-1983,Obukhov:1987yu} in the context of general Einstein-Cartan spacetime where the non-trivial modification of spacetime involving torsion (and hence the spacetime curvature) was introduced sourcing the spin properties of matter. 
In the framework of general relativity, the Lagrangian formulation was developed by Obukhov and Piskareva in \cite{ObukhovPiskareva1989}, where they demonstrated, under the pole-dipole approximation, the conservation laws of the ideal Weyssenhoff spin fluid composed of uncharged particles conform to the general relativistic generalization of evolution equations of spinning test particle (extended body) formulated given by Mathisson \cite{Mathisson:1937zz}, Papapetrou \cite{1951RSPSA.209..248P} and Dixon \cite{1970RSPSA.314..499D}, commonly known as Mathisson-Papapetrou-Dixon (MPD) equations  and the standard continuity equation of an ideal fluid. Eventually a SSC closes the system of MPD equations.\\
\indent
Analogous to the case of spinning test particles, the presence of additional spin-curvature coupling in the momentum balance equation of Weyssenhoff spin fluid presents an interesting avenue for studying the spin and spacetime curvature effects on the morphology of a torus through the spin-curvature coupling in a BH background. With this motivation, for the first time authors of \cite{Lahiri:2023mwj} showed the impacts of Weyssenhoff spin fluid on the morphology of a geometrically thick equilibrium torus.
Assuming a stationary and axisymmetric background with the ideal Weyssenhoff fluid moving in circular motion, a new set of integrability conditions by taking into account of the spin of the fluid, were obtained analogous to relativistic von-Zeipel conditions of an ideal fluid. 
The equilibrium stationary solutions was obtained first in the simplest Schwarzschild spacetime, using the integrability conditons, that allowed the computations of sequences of equilibrium configurations dependent on the fluid spin and the (constant) specific angular momentum distribution in the disk \cite{Lahiri:2023mwj}.\\
\indent In the present paper, we take a step forward and consider an equilibrium torus composed of an ideal Weyssenhoff spin fluid around a Kerr BH with a motivation to incorporate the rotation of the background geometry and examine the morphological changes due to cumulative effects of spin-curvature coupling term in a stationary axisymmetric spacetime represented by Kerr metric. For the purpose of simplicity, we have restricted the present study to a non-self-gravitating torus endowed with constant specific orbital angular momentum distributions.
Furthermore, we consider the scenario in which the spin fluid completely fills its Roche Lobe and take into consideration of the previously obtained integrabilitity conditions (ref.~\cite{Lahiri:2023mwj}) for determining the spin length function, also known as spin density function (to be defined in Section \ref{sec2}). We have computed the stationary equilibrium solutions of the energy density of spin fluid torus in the Kerr background and determined the allowed parameter spaces of the solutions. 

 \textit{Notations and conventions} 
 The Kerr metric is taken with the signature $-,+,+,+$. Geometrized units ($G = c = 1$) will be used throughout the paper. We note that the Greek indices that run over $t,r,\theta,\phi$ denote the standard spherical Boyer-Lindquist coordinates in the Kerr spacetime.

\section{Theoretical framework} 
\label{sec2}

To construct stationary equilibrium solutions of accretion tori comprised of an ideal spin fluid, the torus matter, in the continuum limit is described by the Weyssenhoff ideal neutral spin fluid~\cite{Weyssenhoff:1947iua} whose constituent neutral particles are characterized by the intrinsic spin (whose origin can be attributed to a quantum field theory) in addition to the orbital angular momentum.
The symmetric energy momentum tensor of the Weyssenhoff spin fluid model is given by~\cite{ObukhovPiskareva1989}, 
\begin{equation}
T_{\mu\nu} = (\epsilon+p) u_\mu u_\nu + p g_{\mu \nu} + 2(g^{\rho\sigma}-u^{\rho}u^{\sigma}) \nabla_{\rho} [u_{(\mu}S_{\nu)\sigma}] \, , 
    \label{EM1}
\end{equation}
where $\rho$ is the fluid rest-mass density, $p$ is the fluid pressure, and the four-velocity $u^\mu$ of the fluid is normalized as $u^\mu u_\mu = -1$.
Note that the round brackets in \eqref{EM1} denote the symmetrization of $\mu$ and $\nu$ indices.  The projection tensor defined with respect to the particle's rest frame is given by $\Delta^{\mu\nu} = g^{\mu\nu} + u^\mu u^\nu$. The antisymmetric spin tensor $S^{\mu\nu}$ corresponds to the dipole contribution in the context of the multipole moment expansion~\cite{Weyssenhoff:1947iua}. 
The Frenkel SSC adopted for Weyssenhoff spin fluid is given by,
\begin{equation}
S^{\mu\nu} u_{\nu} =0 \, .  \label{Frenkel}
\end{equation}
where $u^\mu$ defines the four-velocity of the constituent elements of the fluid. 
The divergence of \eqref{EM1} after employing  \eqref{Frenkel} results into the following momentum balance equation,
\begin{equation}
(\epsilon + p) a_\mu + \partial_\mu p + 2 \nabla_\rho (u^\rho S_{\mu\sigma} a^\sigma) + R_{\rho\sigma\tau\mu} S^{\rho\sigma} u^\tau = 0 \label{E1}
\end{equation}
where $R^\mu{}_{\nu\rho\delta}$ is the Riemann curvature tensor and $a^\mu = Du^{\mu}$ is the 4-acceleration (with $D=u^{\alpha}\nabla_{\alpha}$).  It is noted that ~\eqref{E1} represents the general relativistic generalization of the MPD equation under pole-dipole approximation in the context of ideal Weyssenhoff spin fluid model~\cite{ObukhovPiskareva1989} and the spin contribution of the fluid is manifested by its coupling with the spacetime curvature of the background geometry.
The spin four-vector is defined to be,
\begin{equation}
    S_{\mu} = -\frac12 \epsilon_{\mu\nu\rho \sigma} u^{\nu} S^{\rho\sigma} \label{sv}
\end{equation}
and $S^{\mu\nu}$ is given by,
\begin{equation}
    S^{\mu\nu} = -\epsilon^{\mu\nu\rho \sigma} S_{\rho}u_{\sigma} \label{StensorfromSvector}
\end{equation}
The spin density function $S$ is a scalar quantity defined as,
\begin{equation}
  S^2= \frac{1}{2}S_{\mu\nu} S^{\mu\nu} \, . \label{SpinScalarDensty}
\end{equation}
where $S$ can be either positive or negative depending on the orientation of the spin and remains conserved under the Frenkel condition.
The momentum balance equation is further simplified by using the divergence of $S^{\mu}{}_{\rho\sigma} = u^\mu S_{\rho\sigma}$. Using the divergence condition of $S^{\mu}{}_{\rho\sigma}$, and
using the relations $u_{\sigma}a^{\sigma}=0$ and $S^{\lambda \beta}a_{\lambda}a_{\beta}=0$ \cite{Lahiri:2023mwj}, \eqref{E1} gets simplified as follows,
\begin{equation}
  (\epsilon+p) a_\mu + \partial_\mu p +2S_{\mu \beta} Da^{\beta} + R_{\rho\sigma\tau\mu} S^{\rho\sigma} u^\tau =0.  \label{mom-eqn}
\end{equation}
which is the master equation to be solved for obtaining equilibrium solutions of the spin fluid tori under a series of the assumptions which will be discussed in rest of the section. 

The background BH spacetime is stationary and axially symmetric characterised by two Killing vectors namely,
\begin{equation}
    \eta^\mu = \partial_t = (1,0,0,0), \qquad  \xi^\mu = \partial_\phi = (0,0,0,1).
\end{equation} 
resulting into vanishing of the respective Lie derivatives ${\cal L}$ along $\xi$ and $\eta$ of all geometric quantities.
Assuming that the Weyssenhoff fluid shares the same symmetries of the background geometry, any flow parameter $f$ (including the spin) satisfies the following conditions,
\begin{eqnarray}
    {\cal L}_\eta f = 0, \qquad {\cal L}_\xi f = 0,
\end{eqnarray}
implying all flow quantities will only depend on $r$ and $\theta$ in our choice of coordinate system. Given that the Weyssenhoff spin fluid undergoes circular motion around the BH, one can safely turn-off the radial and polar components of the fluid 4-velocity i,e. $u^{r}=u^{\theta}=0$, so that it can be expressed in terms of Killing vectors is written as,
\begin{eqnarray}
u^\mu = A (\eta^\mu + \Omega \xi^\mu) \label{4vel} 
\end{eqnarray}
with the angular velocity defined as,
\begin{eqnarray}
\Omega = \frac{d\phi}{dt} = \frac{u^\phi}{u^t}.
\end{eqnarray}
where $u^{t}$ and $u^{\phi}$ depend on the radial and polar coordinates only.  
It is important to define the dimensionless quantity $l(r,\theta)$ known as the specific orbital angular momentum pertinent to the construction of stationary solutions of a spin fluid torus, which is read off as,
\begin{equation}
- l = \frac{p_\phi}{p_t} = \frac{g_{\phi \mu} u^\mu}{g_{t\nu} u^\nu} = \frac{g_{\phi t} u^t + g_{\phi\phi} u^\phi}{g_{tt} u^t + g_{t\phi} u^\phi} \,. 
\end{equation}
Then from $u_{\mu}=g_{\mu \nu}u^{\nu}$ both $l$ and $\Omega$ are related in the following way,
\begin{equation}
l = - \frac{g_{\phi t} + g_{\phi\phi} \Omega}{g_{tt} + g_{t\phi} \Omega}  \, , \qquad  \Omega = - \frac{g_{t\phi} + g_{tt} l}{g_{\phi\phi} + g_{t\phi} l} \, .
\end{equation}
Note that the normalisation constant $A$ in \eqref{4vel}  can be expressed in terms of either $l$ or $\Omega$ as follows,
\begin{eqnarray}
A = u^t & = & \frac{1}{\sqrt{g_{tt} + 2 g_{t\phi} \Omega + g_{\phi\phi} \Omega^2}} \nonumber\\ 
-u_{t} & = & \displaystyle\sqrt{\frac{g_{t\phi}^2-g_{tt}g_{\phi \phi}}{l^2g_{tt}+2lg_{t \phi}+g_{\phi \phi}}} \nonumber\, .
\end{eqnarray}

The macroscopic spin four-vector $S^\nu$ is always aligned perpendicular to the equatorial plane of the Kerr BH at the equatorial plane, which is given by,
\begin{equation}
    S^\nu = S^\theta \delta^\nu_\theta \, .  \label{spin:1}
\end{equation}
In other words, $S^\nu$ is polar.
Then for the Frankel SSC, the non-zero components of the spin tensor calculated from (\ref{StensorfromSvector}) are as follows,
\begin{eqnarray}
    S^{tr}  =  - S\sqrt{\frac{g_{\theta\theta}}{-g}} u_\phi,\qquad  
  S^{r\phi}  =  -S\sqrt{\frac{g_{\theta \theta}}{-g}}u_{t} \label{compspin1}
\end{eqnarray}
where $S= \sqrt{g_{\theta \theta}}S^{\theta}$ is the spin scalar density.
Due to the symmetries, $S$ is independent of $t$ and $\phi$ in our co-ordinate choice and further remains constant along each particle trajectory for the Frenkel SSC. 
With the symmetry assumptions and under the circular motion of the fluid in a stationary and axially symmetric spacetime, the integrability conditions of momentum conservation equation, presented in~\cite{Lahiri:2023mwj}, have been utilised for determining the solution of spin length function $S$ as well as equilibrium energy density solutions of Weyssenhoff spin fluid tori under a specific choice of specific angular momentum distributions.  For completeness, the integrability conditions associated to \eqref{mom-eqn} are stated below,
\begin{flalign}
\epsilon =  \epsilon(p), \qquad 
\Omega =  \Omega(l)\nonumber  \\
\Phi =  \Phi\left(\frac{A}{\epsilon + p}\right), \quad
\phi  =  \phi\left(\frac{\Omega A}{\epsilon + p}\right), \quad
\partial_{\mu} B= \frac{2 S_{\mu \alpha}\Omega^{\alpha \beta}a_{\beta}}{\epsilon+p}  \label{B-form}
\end{flalign}
the integral form of~\eqref{mom-eqn} becomes, 
\begin{eqnarray}
W&-&W_{\mathrm{in}}+\int_{0}^{p}\frac{dp}{\epsilon+p} \\ \nonumber
&+&\int_{0}^{r_{\mathrm{in}}} d B+\int_{0}^{\Phi}\frac{A(r,\theta)d\Phi}{\epsilon+p}+
\int_{0}^{\phi}\frac{\Omega A(r,\theta)d\phi}{\epsilon+p}=0 \label{Potential}
\end{eqnarray}
with the effective potential,
\begin{eqnarray}
W-W_{\mathrm{in}} = \ln{|u_{t}|}- \ln{|u_{t_{\mathrm{in}}}|}- \int_{l_{\mathrm{in}}}^{l} \frac{\Omega dl}{1-\Omega l}.
\end{eqnarray} 
Here $W_{\mathrm{in}}$ is the total potential at the inner edge of the torus at the equatorial plane.
$\Omega_{\alpha \beta}$ is the antisymmetric angular velocity tensor (see, for details~\cite{Lahiri:2023mwj}). Note that~\eqref{Potential} is obtained by considering both the pressure and the energy density of the spin fluid to vanish at the outer surface and on the inner edge of the torus.  Also at the radial position of inner edge i,e. $r=r_{\mathrm{in}}$, one usually has $l=l_{in}$, $u_{t}= u_{t_{in}}$ and $\Phi_{in}=0=\phi_{in}$.  
It is easily seen that \eqref{B-form} becomes the relativistic von-Zeipel condition in absence of macroscopic spin and \eqref{Potential} reduces to the usual hydrostatic equilibrium condition of an ideal fluid without spin.

\begin{figure*}[htb!]
\captionsetup{justification=raggedright, singlelinecheck=on}
\hspace{-2.6cm}
\begin{subfigure}[b]{0.4\textwidth}
  \includegraphics[scale=0.6]{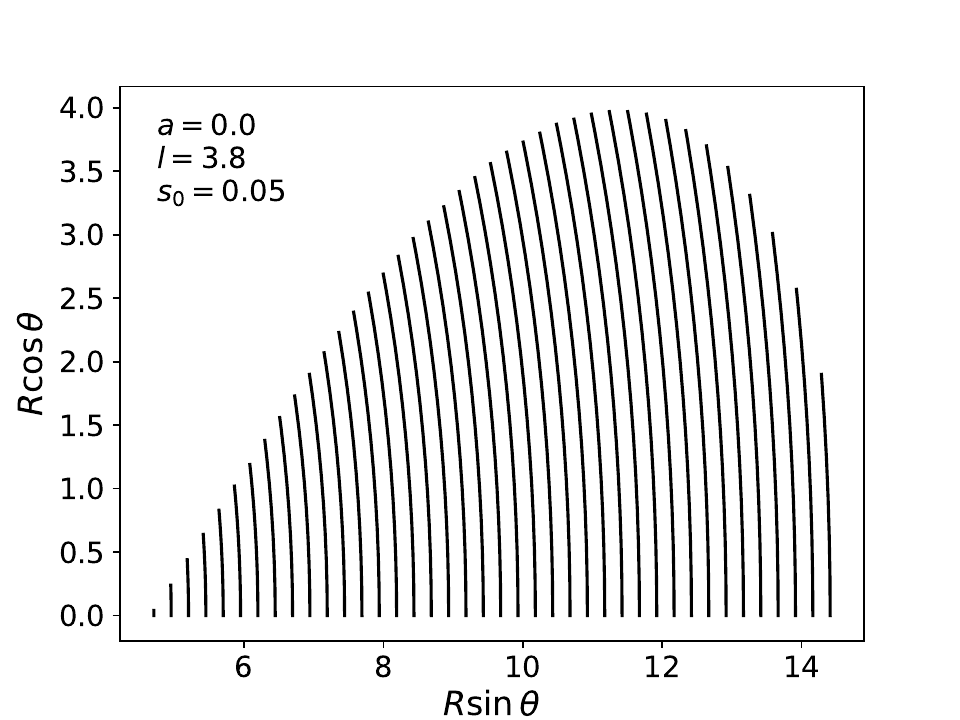}
    \end{subfigure}
    \hspace{2.3cm}
  \begin{subfigure}[b]{0.4\textwidth}
   \includegraphics[scale=0.6]{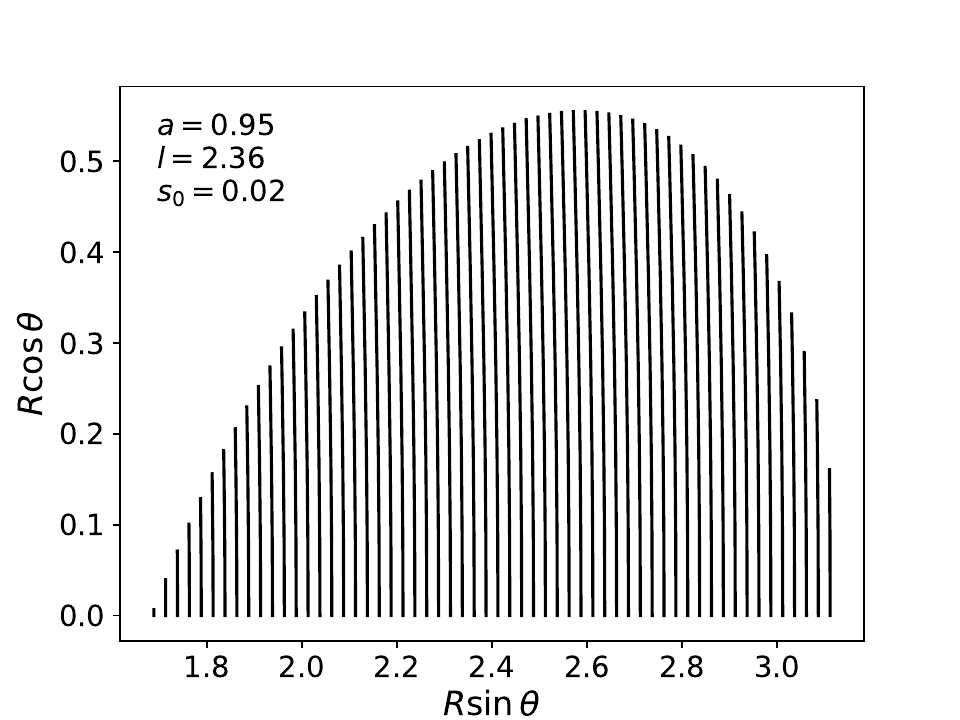}
    \end{subfigure}
   \caption{Characteristic curve structure of $k(r, \theta)$ for a Schwarzschild BH with an orbital angular momentum of $l=3.8$ and $s_0 = 0.05$ (left panel) and for a Kerr BH with $a = 0.95$, $l = 2.36$ and $s_0 = 0.02$ (right panel). Notice that the number of characteristic curves for each model has been sampled down to facilitate the visualization of their structure.}  \label{fig-char}
   \end{figure*}

\section{Methodology}
In order to construct stationary equilibrium models of Weyssenhoff spin fluid tori in a rotating background, we start from simplest rotating BHs described by the Kerr metric. We consider the torus in the hydrostatic equilibrium is endowed with constant specific orbital angular momentum distributions. The macroscopic spin vector of the fluid is always aligned perpendicular to the equatorial plane of the Kerr BH which also is assumed to coincide with the equatorial plane of the torus. The spin fluid, which obeys the same symmetries of the Kerr spacetime, undergoes circular motion around the Kerr BH. We additionally assume the internal energy density of the fluid is very small, as a result, the total energy density is approximately equal to the rest-mass density. 
\subsection{Formalism}
The Kerr metric, parametrized by the mass parameter $M$ and the spin parameter $a$ in the standard Boyer-Lindquist coordinates reads as,
\begin{equation}
    ds^2 = g_{tt}dt^2 + 2 g_{t\phi}dt d\phi  + g_{rr} dr^2+
    g_{\theta \theta} d\theta^2 + g_{\phi \phi} d\phi^2
\end{equation}
with the metric components being,
\begin{eqnarray}
g_{tt}&=& -\left(1-\frac{2Mr}{\Sigma(r,\theta)}\right),\qquad  g_{rr} =\frac{\Sigma(r,\theta)}{\Delta(r)},\qquad
g_{\theta \theta} =\Sigma(r,\theta) \nonumber \\
g_{t\phi} &=& -\frac{2Mra\sin^2\theta}{\Sigma(r,\theta)}, \quad
g_{\phi \phi}=\frac{(\varpi^2-a^2\Delta \sin^2\theta)\sin^2\theta }{\Sigma(r,\theta)} 
\end{eqnarray}
where, 
\begin{eqnarray}
    \Sigma(r,\theta)& =& r^2 +a^2 \cos^2 \theta,\nonumber \\
    \Delta(r) = (r^2 +a^2)-2Mr, \quad
    \varpi^2 &=& (r^2 +a^2)^2 \nonumber 
\end{eqnarray}
Without any loss of generality, we set $M=1$ in this work.
The specific orbital angular momentum distribution is set as $l(r, \theta)=l_0$ where $l_0$ is a constant. Considering a polytropic equation of state as $p=\kappa \epsilon^{\gamma}$, where $\kappa$ is the polytropic constant and $\gamma$ is the adiabatic index, 
 the momentum balance equation \eqref{mom-eqn} in accordance with the spacetime symmetries reduces to,
\begin{flalign}
    \partial_{\mu} \epsilon =\frac{(\epsilon^{2-\gamma}+\kappa \epsilon)}{-\kappa \gamma}\partial_{\mu} \ln(-u_t{})- \frac{(2S_{\mu \beta}Da^{\beta}+R_{\rho \sigma \tau \mu}S^{\rho \sigma}u^{\tau})}{\kappa \gamma \epsilon^{\gamma-1}} \label{mom-eqn1}
\end{flalign}
Using the non-zero components $S^{tr}$ and $S^{r\phi}$ of the spin tensor given by \eqref{compspin1}, the radial and polar components of \eqref{mom-eqn1}, are expressed as follows,
\begin{eqnarray}
  \partial_{r} \epsilon &=& \frac{(\epsilon^{2-\gamma}+\kappa \epsilon)}{-\kappa \gamma}\partial_{r}\ln (-u_{t}) -\frac{S(r,\theta)}{\kappa \gamma\epsilon^{\gamma-1}(1-\Omega l_0)}\sqrt{\frac{g_{\theta \theta}}{-g}} F_1\nonumber \\
&-&  G_1(r, \theta)\,[(u_{\phi}-\frac{g_{t \phi}}{g_{tt}}u_{t} )Da^{t}+(\frac{g_{\phi t}}{g_{tt}}u_{\phi}-\frac{g_{\phi \phi}}{g_{tt}}u_{t})Da^{\phi}]
\label{den-1}\nonumber \\
 \\[2mm]
\partial_{\theta}\epsilon &=& \frac{(\epsilon^{2-\gamma}+\kappa \epsilon)}{-\kappa \gamma}\partial_{\theta}\ln (-u_{t})-\frac{S(r,\theta)}{\kappa \gamma\epsilon^{\gamma-1}(1-\Omega l_0)}\sqrt{\frac{g_{\theta \theta}}{-g}}F_2
\nonumber \\ 
\label{den-2} 
\end{eqnarray}
where $G_1(r,\theta)=  \displaystyle\frac{2g_{rr}\,g_{tt}\,S(r,\theta)}{\kappa \gamma\epsilon^{\gamma-1}} \sqrt{\frac{g_{\theta \theta}}{-g}}$.
Using the following relations,  
\begin{eqnarray}
u_{\phi}u^{t} &=& \frac{l}{1-\Omega l_0},\quad 
u_{t}u^{\phi}=-\frac{\Omega}{1-\Omega l_0}. \label{relations}
\end{eqnarray}
the functions $F_1$ and $F_2$ in \eqref{den-1} and \eqref{den-2} are given by,
\begin{eqnarray}
F_1(r,\theta)&=& - \Omega l_0 R_{tr\phi r}+\Omega R_{r \phi \phi r}- l_0 R_{trtr}+R_{r \phi t r}\,, \nonumber \\
F_2(r,\theta)&=& - \Omega l_0 R_{tr\phi \theta }+\Omega R_{r \phi \phi \theta}- l_0 R_{tr t\theta}+ R_{r \phi t \theta}
\end{eqnarray}

Note that in contrary to the Schwarzschild case, in Kerr spacetime, both $F_1$ and $F_2$ are non-zero due to additional non-zero components of Riemann curvature tensor.  
The equilibrium solutions are then constructed by computing the energy density and subsequently the fluid pressure using the set of first-order partial differential equation~\eqref{den-1} and~\eqref{den-2} for a given spin length function $S$ together with the values of $\kappa$ and $\gamma$. 
As the spin length function varies only with respect to radial and polar coordinates, following ~\cite{Lahiri:2023mwj} let us consider the following ansatz for $S$ as,
\begin{eqnarray}
    S(r,\theta) = s_0 \epsilon^{\gamma-1}k(r,\theta)(1-\Omega l_0) \label{eq:15}
\end{eqnarray}
where $s_0$ is the constant spin parameter associated with the overall intrinsic spin of the fluid elements and $l_0$ is the constant value of the specific orbital angular momentum of the fluid. Note that $s_0$ can be both positive or negative depending on the alignment of the spin vector about the equatorial plane of the Kerr BH. 
In general, $aS>0$ corresponds to spin of the fluid is parallel to the spin of the Kerr BH whereas $aS<0$ corresponds to the antiparallel combination. 

In what follows, though considered to be high for astrophysical systems, taking $\gamma=2$ leads to a simple and tractable forms of the radial and the polar components of the momentum balance equation. Hence using ~\eqref{eq:15} with $\gamma=2$,  \eqref{den-1} and ~\eqref{den-2} reduce to,
\begin{eqnarray}
  \partial_{r} \epsilon &=& \frac{(1+\kappa \epsilon)}{-2\kappa}\partial_{r}\ln (-u_{t}) -\frac{s_0 k(r,\theta)}{2\kappa}\sqrt{\frac{g_{\theta \theta}}{-g}} F_1\nonumber \\
&-&  G_1(r, \theta)\,[(u_{\phi}-\frac{g_{t \phi}}{g_{tt}}u_{t} )Da^{t}+(\frac{g_{\phi t}}{g_{tt}}u_{\phi}-\frac{g_{\phi \phi}}{g_{tt}}u_{t})Da^{\phi}]
\label{den-3}\nonumber \\
 \\[2mm]
\partial_{\theta}\epsilon &=& \frac{(1+\kappa \epsilon)}{-2\kappa}\partial_{\theta}\ln (-u_{t})-\frac{s_0 k(r,\theta)}{2\kappa}\sqrt{\frac{g_{\theta \theta}}{-g}}F_2
\nonumber \\ 
\label{den-4} 
\end{eqnarray}
where $G_1(r,\theta)=\displaystyle\frac{2s_0g_{rr}\,g_{tt}\,k(r,\theta)(1-\Omega l_0)}{2\kappa}$ in this setup.\\
The existence of integrability conditions clearly implies $\frac{dp}{p+\epsilon}$ is an exact differential. Additionally, they imply the energy density and the pressure are related by an equation of state $\epsilon=\epsilon(p)$.  Therefore for determining the function $k(r,\theta)$ in \eqref{eq:15}, we apply of the compatibility condition of the energy density $\epsilon$. The simplicity that comes with the choice $\gamma = 2$ is now more apparent, as it can be seen that the PDE for $k(r, \theta)$ that comes from the compatibility condition involving the second derivatives of $\epsilon$ will be independent of $\epsilon$ thereby effectively decoupling the magnitude of $k(r, \theta)$ from $\epsilon$ (but not the other way around). By using the obtained solution of $k(r, \theta)$, $\epsilon(r,\theta)$ (and hence the pressure $p=\kappa \epsilon^2$) is determined by numerically solving \eqref{den-3} and \eqref{den-4}. Subsequently the spin length function is determined from \eqref{eq:15}.   This process will be made self-evident in the following section where the description on the applied numerical techniques and solution-obtaining procedure are discussed in detail.

Generally, a closed equilibrium torus is characterised by the radial locations of its cusp, center, inner edge and the outer edge respectively denoted by $r_{\mathrm{cusp}}, r_{\mathrm{max}}, r_{\mathrm{in}}$ and $r_{\mathrm{out}}$ all defined at the equatorial plane. The cusp and the center are defined as geodesic ($a_{\mu} = 0$) points (curves). The cusp corresponds to a saddle point whereas the maximum for the rest-mass density (or, equivalently fluid pressure or energy density) is associated to the center of the torus. Moreover the inner edge can be chosen at any point between the cusp and the center whereas the outer edge of the torus is fixed as the location of the inner edge is chosen.
Here we restrict to the case of a critically-filled torus for which the effective potential gap $\Delta W$ satisfies $\Delta W=W_{\mathrm{in}}-W_{\mathrm{cusp}}=0$ indicationg the inner edge of the torus exactly coincides with its cusp ($r_{\mathrm{in}} = r_{\mathrm{cusp}}$) and thus corresponds to a torus critically filling its "Roche lobe". This further implies that at the inner edge, both the energy density as well as its first derivative exactly vanish. 
Given that in a stationary and axially symmetric spacetime, ISCOs of spinning particles get modified due to spin-curvature coupling (see~\cite{Jefremov:2015gza} and references therein), and that the Weyssenhoff fluid allows spin-curvature coupling as manifested  by the momentum balance equation, it is therefore natural to expect that the characterised radii, namely, the radius of marginally stable circular orbit $r_{\mathrm{ms}}$ and the radius of bound circular orbit $r_{\mathrm{mb}}$ of a circularly rotating ideal spin fluid in the Kerr spacetime must be modified in presence of spin and the corresponding curvature coupling with spin.  Additionally, as the magnitude of the Keplerian angular momentum $l_{k}$ (determined from the condition $a_{\mu} =0$) gets altered due to macroscopic spin of the fluid, it further implies deviations of specific angular momentum $l_{\mathrm{mb}}$ and $l_{\mathrm{ms}}$ defined respectively at $r_{\mathrm{mb}}$ and $r_{\mathrm{ms}}$ from the exact counterparts in absence of spin in Kerr spacetime.
For obtaining closed tori solutions, we will restrict the constant value of $l_0$ between $l_{\mathrm{mb}}^{\mathrm{Kerr}}$ and $l_{\mathrm{ms}}^{\mathrm{Kerr}}$ for a given Kerr spin parameter $a$ and $l_0$.  Also, we have set $\kappa=1$ throughout this work. 
\subsection{Numerical techniques}
To solve Eqs.~\eqref{den-3} and~\eqref{den-4} we go through the following procedure:
In the first place, we obtain an extra equation for the function $k(r, \theta)$ by using the compatibility condition of the mixed second derivatives of the energy density $\partial_{\theta}\partial_{r}\epsilon(r, \theta) - \partial_{r}\partial_{\theta}\epsilon(r, \theta)$ and substitute Eqs.~\eqref{den-3} and~\eqref{den-4} again in the expression we obtain the following PDE for $k(r, \theta)$
\begin{eqnarray}\label{eq:k_eq}
F^{*}_{2} (\partial_r k) - (F^{*}_1 + G^{*})(\partial_{\theta} k ) \,+ k(\partial_{r} F^{*}_2  - \partial_{\theta} F^{*}_1 - \partial_{\theta} G^{*} )\nonumber  \\ 
+\frac{k}{2} \left(-F^{*}_{2} \partial_r \ln (-u_t) + (F^{*}_1 + G^{*})\partial_{\theta} \ln (-u_t)\right)   = 0\,
\end{eqnarray}
where the functions $F^{*}_{1}$, $F^{*}_{2}$ and $G^{*}$ are only functions of the radial and the polar coordinates and are defined as
\begin{eqnarray}
  F^{*}_{1} &=& \frac{s_0}{2\kappa}\sqrt{\frac{g_{\theta \theta}}{-g}} F_1, \quad   F^{*}_{2} = \frac{s_0}{2\kappa}\sqrt{\frac{g_{\theta \theta}}{-g}} F_2 \\
[2mm]
  G^{*} &=& \frac{2s_0g_{rr}\,g_{tt}\,(1-\Omega l_0)}{2\kappa}[(u_{\phi}-\frac{g_{t \phi}}{g_{tt}}u_{t} )Da^{t} \nonumber \\ 
  &+& (\frac{g_{\phi t}}{g_{tt}}u_{\phi}-\frac{g_{\phi \phi}}{g_{tt}}u_{t})Da^{\phi}] 
\end{eqnarray}
We further compatify~\eqref{eq:k_eq} to express it as
\begin{equation}\label{eq:k_simple}
A_k \partial_r k + B_k \partial_{\theta} k + C_k k = 0\,,
\end{equation}
where the coefficients $A_k$, $B_k$ and $C_k$ are just the coefficients in~\eqref{eq:k_eq} for the radial, polar and independent terms respectively.
Noticing that all the coefficients of~\eqref{eq:k_simple} are only functions of $(r, \theta)$ gives us the capability of solving for $k(r, \theta)$ irrespective of the values of $\epsilon$, therefore simplyfing the procedure. Taking now the limit $\theta = \pi/2$ gives us the following equation at the equatorial plane
\begin{equation}
B_k|_{\theta = \pi/2} \partial_{\theta} k = 0\,,
\end{equation}
whith $B_k|_{\theta = \pi/2}$ different from 0. This fact gives us two pieces of information. The first is that~\eqref{eq:k_eq} does not directly give information about $k$ in the equatorial plane, and the second is that the $k$ constant curves are going to be perpendicular to the equatorial plane. Further inspecting~\eqref{eq:k_eq} tells us that, because of the fact that the indepedent term is directly proportional to $k$ and the coefficients $A_k$ and $B_k$ are independent of $k$, i) if $k$ vanishes at the equatorial plane, it will vanish everywhere (i.e.~$k(r, \pi/2) = 0 \implies k(r, \theta) = 0$) ii) it is possible to rescale the value of $k$ at any point to $1$, reabsorbing the extra factor in the paramater $s_0$, this means that the shape of the characteristic curves of the function $k(r,\theta)$ do not depend on the value of $k(r, \theta)$ and iii) a constant $k$ solution is not possible for $k\neq0$. For the sake of comparison with the Schwarzschild case, we are going to choose $k(r, \pi/2) = 1$ as the initial condition for the characteristic curves at the equatorial plane. Starting from that, we solve Eq.~\eqref{eq:k_simple} numerically using the method of characteristics to find the distribution of $k(r, \theta)$ in our region of interest. In Figure~\ref{fig-char} we can see an example of the characteristic structure of the function $k(r, \theta)$ for a Schwarzschild BH with $l = 3.8$ and $s_0 = 0.05$ and for a Kerr BH with $a = 0.95$, $l= 2.36$ and $s_0 = 0.02$. The numerical integration is performed using the standard Runge-Kutta 4th order method.\\

  In the second place, now that we have $k(r, \theta)$ we are able to solve~Eqs.~\eqref{den-3} and~\eqref{den-4}. First, the location of the cusp $r_{\mathrm{cusp}}$ is obtained by evaluating~Eq.~\eqref{den-3} at the cusp (which implies $\partial_{r} \epsilon = \epsilon = 0$ according to our assumptions) This gives us the following expression that we can solve for $r$ to obtain the value of $r_{\mathrm{cusp}}$
  \begin{eqnarray}
  0 &=& \frac{(1)}{-2\kappa}\partial_{r}\ln (-u_{t}) -\frac{s_0 k(r,\pi/2)}{2\kappa}\sqrt{\frac{g_{\theta \theta}}{-g}} F_1\nonumber \\
 &-&  G_1(r, \theta)\,[(u_{\phi}-\frac{g_{t \phi}}{g_{tt}}u_{t} )Da^{t}+(\frac{g_{\phi t}}{g_{tt}}u_{\phi}-\frac{g_{\phi \phi}}{g_{tt}}u_{t})Da^{\phi}]
  \end{eqnarray}
  this equation is solved numerically using the standard Newton method, using the location of the cusp of a spin-less fluid disk as the initial guess. Once we have the location of the cusp, we can integrate~Eq.~\eqref{den-3} numerically starting from the cusp, in which $\epsilon$ vanishes until the point in which it vanishes again (i.e.~$r_{\mathrm{out}}$). In the process, the radial distribution of the energy density at the equatorial plane $\epsilon(r, \pi/2)$ is found, together with the value of $r_{\mathrm{max}}$ and the maximum of the energy density $\epsilon_{\mathrm{max}} = \epsilon ({r_\mathrm{max}})$. And finally, using the energy density radial distribution as initial values, we can integrate numerically Eq.~\eqref{den-4} to obtain a sufficiently dense set of curves that map the entirety of the disk. For both Eq.~\eqref{den-3} and~\eqref{den-4} we use an standard Runge-Kutta 4th order solver.

  It is relevant to mention that, in practice, we solve first for the radial distribution of $\epsilon$ at the equatorial plane, to find the characteristic radii, and the we run a first test integration of $\epsilon$ in the whole plane but using a constant value of $k(r, \theta) = 1$. As it was mentioned before, that does not solve the problem ($k$ constant is not a solution of~\eqref{eq:k_eq}), but it gives us a guess of the vertical height that the disk will have, and therefore it is useful to put limits on the vertical height that the characteristic curves are allowed to achieve as, in some cases, they exist well beyond the extent of the disk. Once the guess is obtained, the procedure is followed as previously described.

\section{Results} \label{res}
The primary motivation of this work is to investigate the impacts of the macroscopic spin of the disk fluid using Weyssenhoff fluid phenomenological model on stationary equilibrium solutions of a non-self gravitating torus located in the Kerr spacetime. As the intrinsic spin of the fluid substrate is manifested by the spin-curvature coupling term, the consequences of the fluid spin on the solutions relies on the properties of the background Kerr metric. Accordingly, we have explored the impacts of spin contributions pertaining to two different origins and their cumulative effects on the equilibrium solutions of the torus. 
\subsection{Morphology of the disks}
\begin{figure*}[htb!]
\captionsetup{justification=raggedright, singlelinecheck=on}
\hspace{-2.6cm}
\begin{subfigure}[b]{0.4\textwidth}
  \includegraphics[scale=0.5]{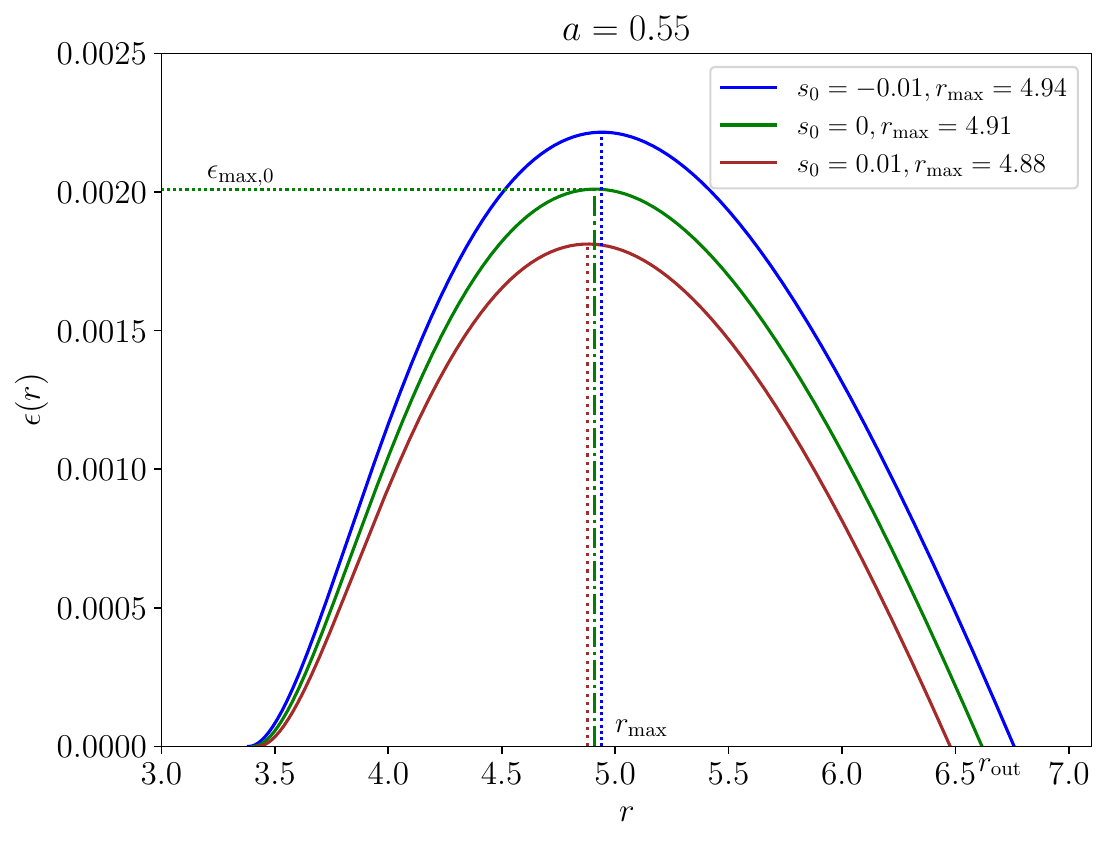}
    \end{subfigure}
    \hspace{2.3cm}
  \begin{subfigure}[b]{0.4\textwidth}
  	\includegraphics[scale=0.5]{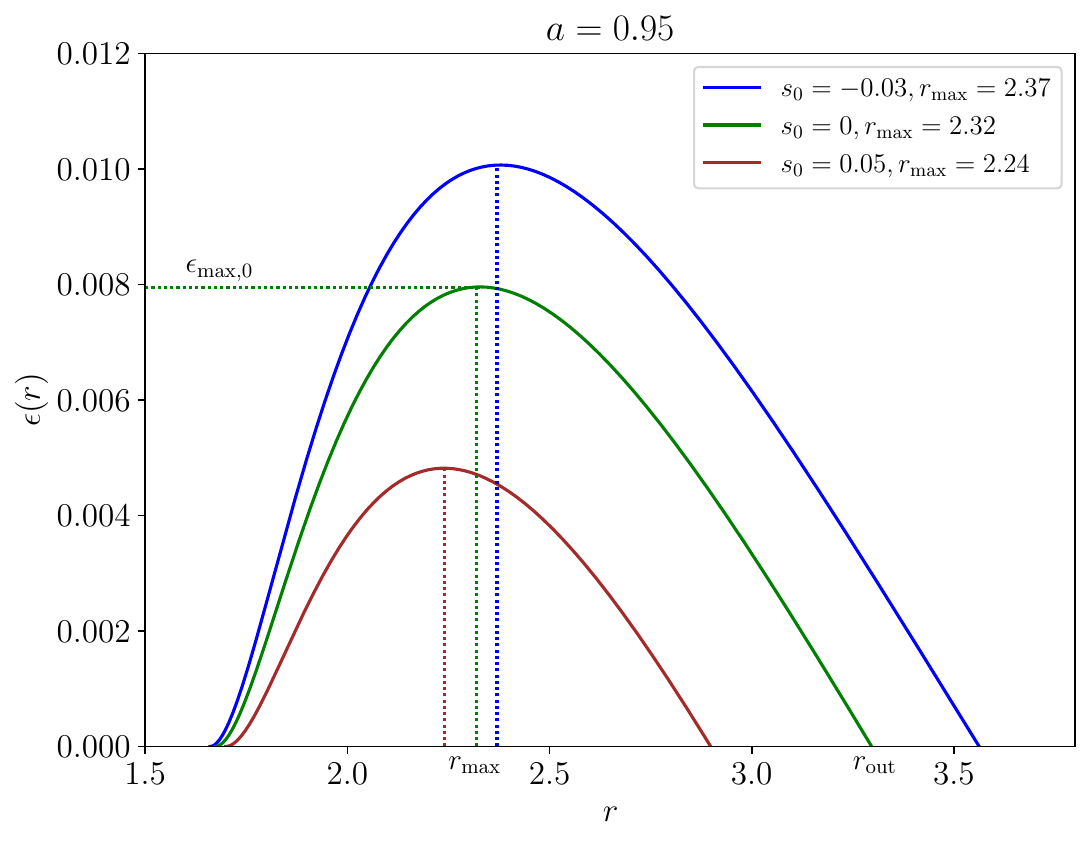}
    \end{subfigure}
  	\caption{Energy density of a torus at the equatorial plane supported by the Weyssenhoff spin fluid for fluid spin parameter $s_0>0$, $s_0<0$ and $s_0=0$ for co-rotating motion. Note that $r_\mathrm{max}$ corresponds to the radial location of the maximum energy density of the torus and $\epsilon_{\mathrm{max},0}$ corresponds to the magnitude of $\epsilon_{\mathrm{max}}$ with $s_0=0$. Two values of Kerr spin parameter are considered here for comparison, namely, in the left panel $a=0.55,l_0=3.14$ and in the right panel $a=0.95, l_0=2.36$ are taken respectively. }  \label{fig-rc}
   \end{figure*}

We now present the stationary equilibrium energy density solutions of a critically-filled closed spin fluid torus characterised by constant orbital angular momentum distributions for both corotating and counter-rotating cases. Here Figure~\ref{fig-rc} presents the radial energy density profiles obtained at the equatorial plane for two different values, namely, $a=0.55$ corresponding to moderately spinning and  $a=0.95$, a highly spinning Kerr BH. Here, $l_0$ for every value of $a$ is chosen by having a small deviation (roughly about $2\%$) from $l_{\mathrm{ms}}^{\mathrm{Kerr}}$ for a given $a$. 
For each value of $a$, three magnitudes of $s_0$ are considered including $s_0=0$ to facilitate a comparison with an equilibrium torus without macroscopic spin.  One immediately notes from Figure~\ref{fig-rc}, the spin fluid torus becomes enlarged for $s_0<0$, while on the other hand, a positive magnitude of $s_0$ turns it smaller in comparison to an ideal torus without any spin. This aspect also shows up from the shift of the outer edge $r_\mathrm{out}$ of the disk with varying $s_0$. 
It is further observed the peak value of total radial energy density of a torus, denoted by $\epsilon(r)$, gets enhanced with $s_0<0$ and diminishes with $s_0>0$ in comparison to the energy density of no-spin fluid torus (denoted by $\epsilon_{\mathrm{max,0}}$), a pattern that prevails irrespective of the magnitude of the rotation of the Kerr BH.  It is also observed that, not only the outer edge, but the radial location of center of the torus i,e. $r_{\mathrm{max}}$ gets relocated in presence of fluid spin when compared to the respective counter-part of an ideal torus without Weyssenhoff fluid. In fact, $r_{\mathrm{max}}$ moves away from the outer horizon of the Kerr BH corresponding to negative $s_0$ values whereas get shifted towards the horizon of a torus with positive magnitudes of $s_0$. 

\begin{figure*}[htb!]
\captionsetup{justification=raggedright, singlelinecheck=on}
\hspace{-2cm}
\begin{subfigure}[b]{0.4\textwidth}
  	\includegraphics[scale=0.5]{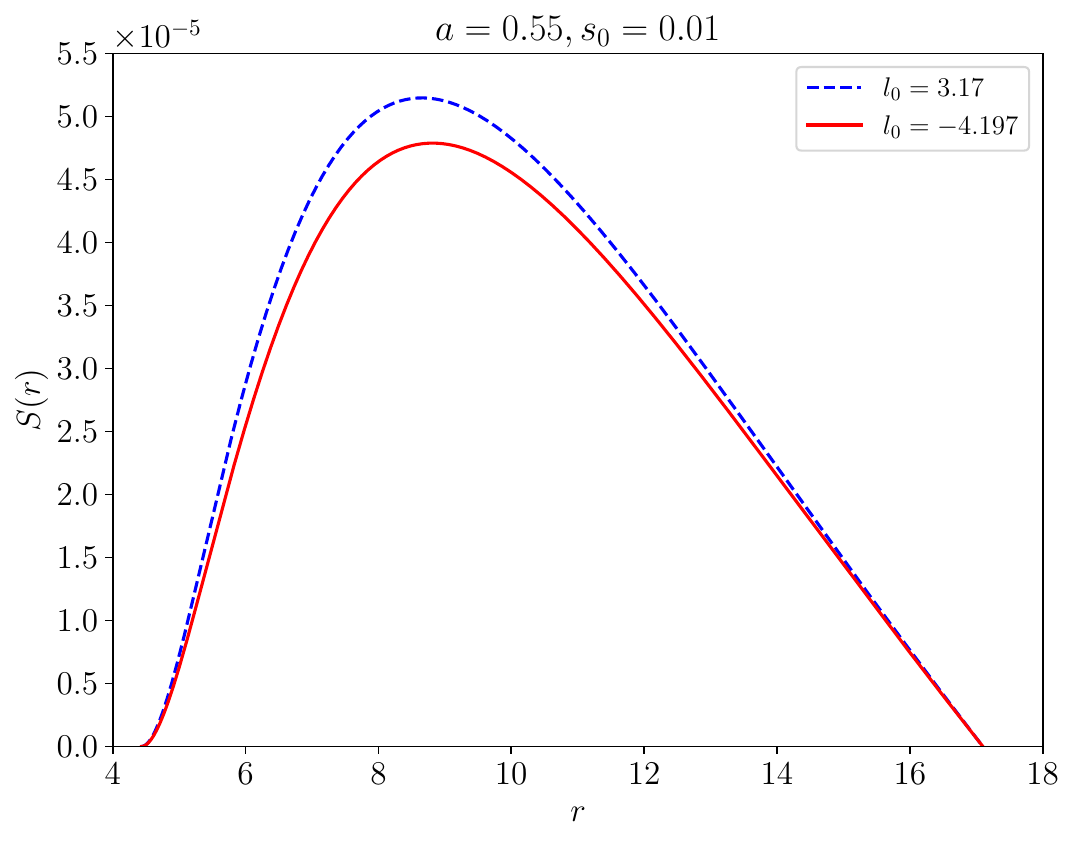}
    \end{subfigure}
    \hspace{2.3cm}
  \begin{subfigure}[b]{0.4\textwidth}
  	\includegraphics[scale=0.5]{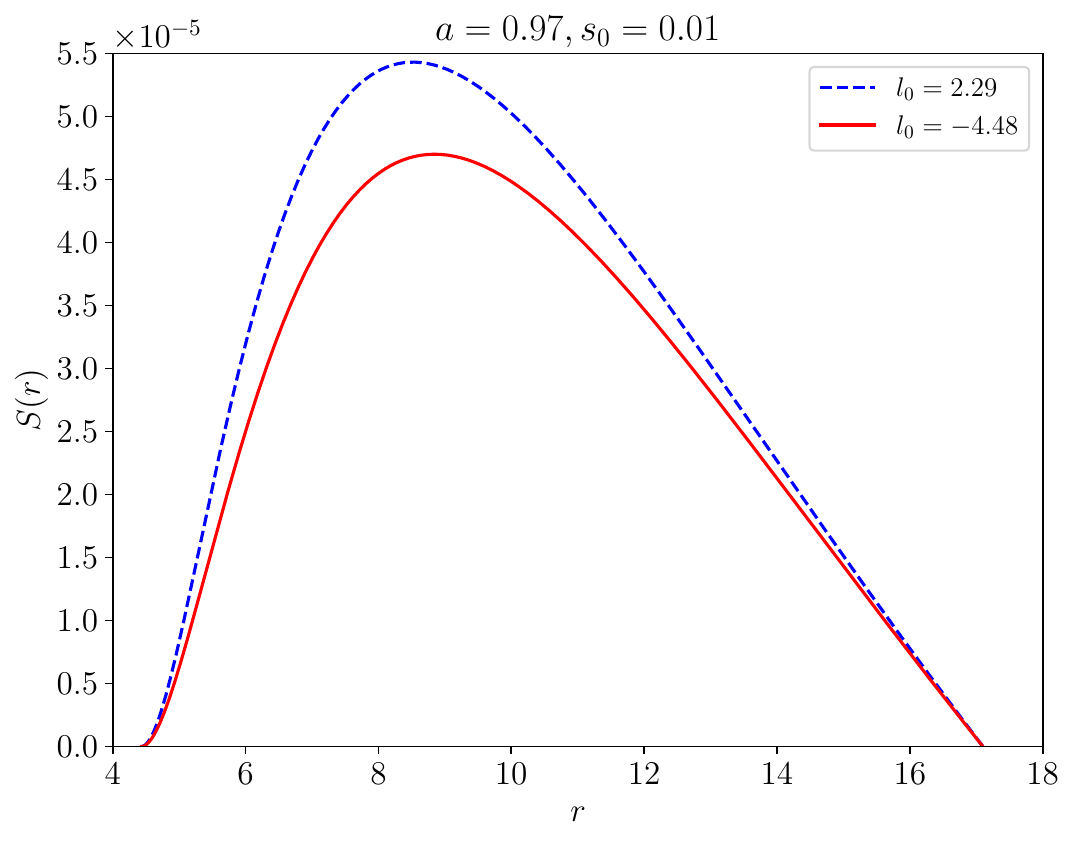}
    \end{subfigure}
   \caption{Behaviour of the spin length function $S(r)$ is depicted at the equatorial plane corresponding to $a_0=0.55$ (left panel) and $a_0=0.97$ (right panel). For simplicity, the same magnitude of spin parameter $s_0=0.01$ is considered in both panels.}  \label{SpinL}
  \end{figure*}

The radial profile of the spin length function determined using \eqref{eq:15} is then presented in Figure~\ref{SpinL} for both corotating ($l_0>0$) and counter-rotating ($l_0<0$) motions. To understand the behaviour of the spin length function, we have typically considered cases of moderately and highly rotating Kerr spacetime, and for the convenience of analysis, fixed the magnitude of the spin parameter $s_0$ of the fluid to $s_0=0.01$. Since $k(r,\pi/2=1)$, the spin density function being proportional to the total energy density takes in higher peak value for corotating (prograde) motion in comparison to the counter-rotating (retrograde) motion. Moreover, the peak positions corresponding to $l_0>0$ and $l_0<0$ never coincide. (However, note that, by fixing $s_0$, the inner and the outer edges essentially overlap irrespective of different $l_0$ in both panels.)

\begin{figure*}[htb!]
\captionsetup{justification=raggedright, singlelinecheck=on}
\hspace{-2cm}
\begin{subfigure}[b]{0.4\textwidth}
  	\includegraphics[scale=0.5]{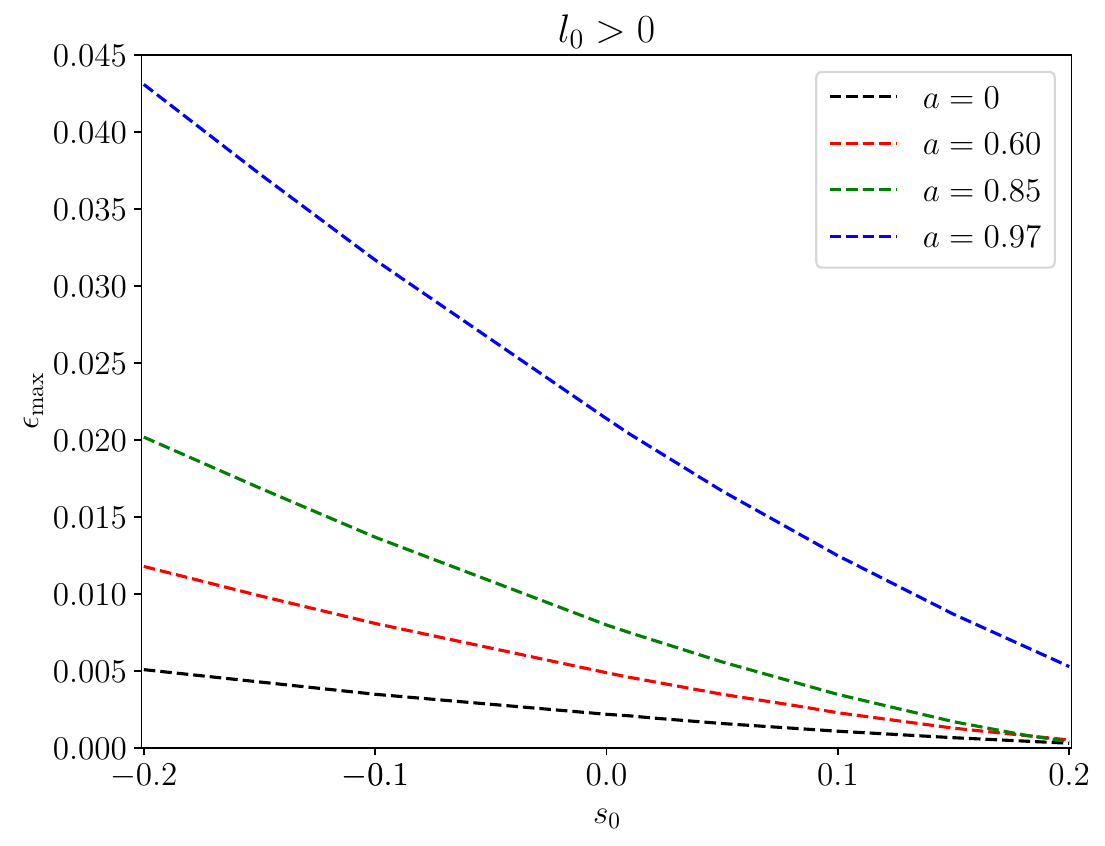}
    \end{subfigure}
    \hspace{2.3cm}
  \begin{subfigure}[b]{0.4\textwidth}
  	\includegraphics[scale=0.5]{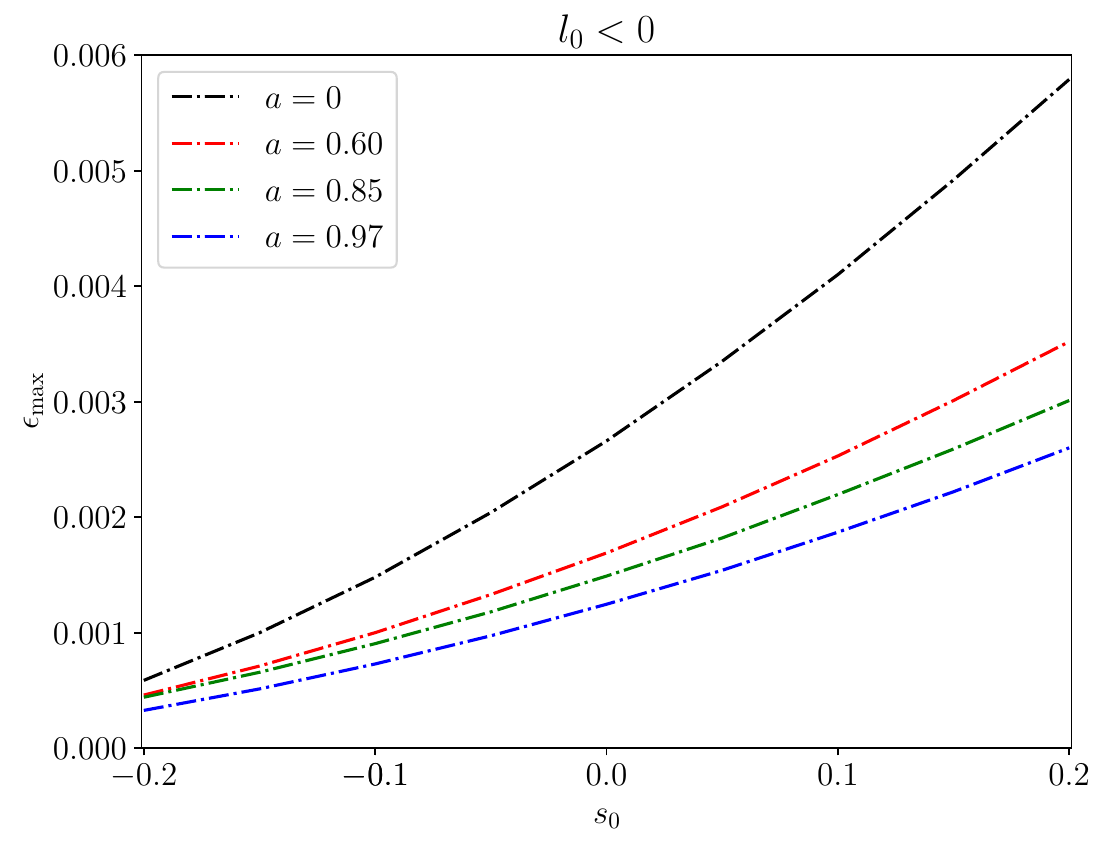}
    \end{subfigure}
   \caption{Behaviour of $\epsilon_{\mathrm{max}}$ with respect to $s_0$ for different values of Kerr spin parameter. The left panel corresponds to the co-rotating case ($l_0>0$) and the right panel depicts the counter-rotating case ($l_0<0$). }  \label{emax-s0}
  \end{figure*}

Following Figure~\ref{fig-rc} which focuses on consequences of fluid spin on equilibrium total density solutions for a fixed value of $a$, in fact, the diminution or enhancement of the equatorial plane energy density with respect to the variation of $s_0$ is rather found to be generic as shown in Figure~\ref{emax-s0} where both positive and negative values of $s_0$ are considered as well as multiple values of $a$ for both corotating (left panel) and counter-rotating (right-panel) motions. In the left panel, it is observed that if one increases from $s_0=0$ to positive values, $\epsilon_{\mathrm{max}}$ rapidly decreases whereas $\epsilon_{\mathrm{max}}$ monotonically increases by decreasing $s_0$ to negative values.
This feature gets completely reversed for counter-rotating motion of the Kerr BH as seen on the right panel where increasing values from $s_0=0$ to positive values increases the magnitudes of $\epsilon_{\mathrm{max}}$ while it decreases as $s_0$ is decreased to negative values. It is also observed that, the profile of $\epsilon_{\mathrm{max}}$ becomes steeper as $a$ is increased to highly rotating Kerr spacetime from $a=0$ (Schwarzschild case) for corotating motion whereas one notes a completely opposite behaviour of $\epsilon_{\mathrm{max}}$ for counter-rotating motion.
It is worthwhile to note the behaviour of $\epsilon_{\mathrm{max}}$ for the co-rotating motion exactly follows the same pattern as observed in the Schwarzshild spacetime \cite{Lahiri:2023mwj}.

\begin{figure*}[htb!]
\captionsetup{justification=raggedright, singlelinecheck=on}
\hspace{-2cm}
\begin{subfigure}[b]{0.4\textwidth}
    \includegraphics[scale=0.5]{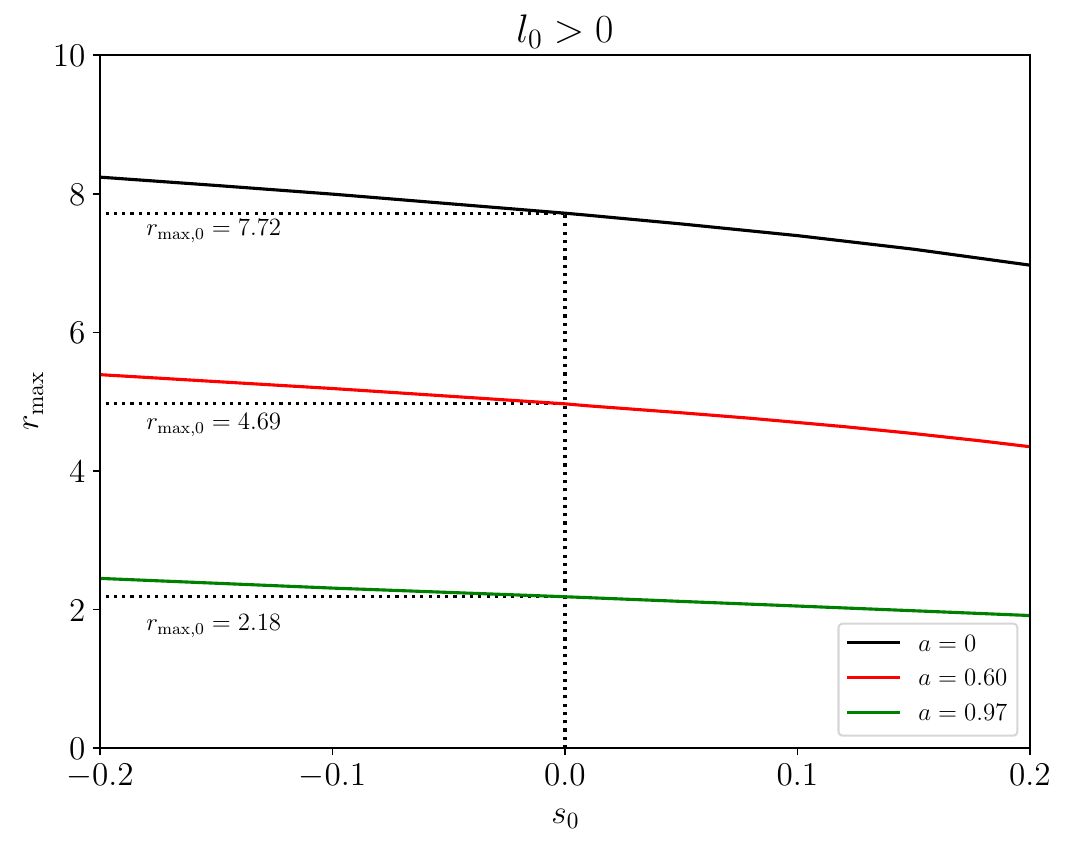}
\end{subfigure}
   \hspace{2cm}
  \begin{subfigure}[b]{0.4\textwidth}
  	\includegraphics[scale=0.5]{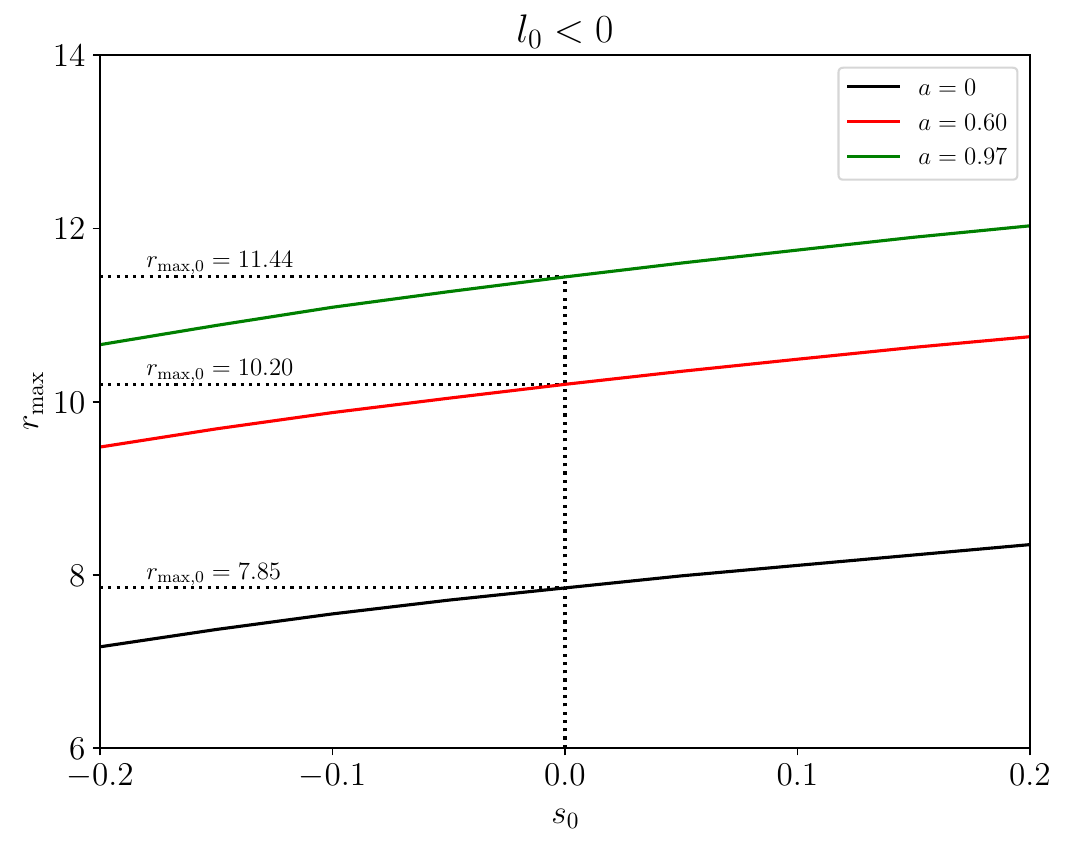}
    \end{subfigure}
    \caption{The behaviour of $r_{\mathrm{max}}$ with respect to $s_0$ is depicted for different values of $a$. The left and the right panels respectively depict the co-rotating and the counter-rotating cases. Here $r_{\mathrm{max},0}$ corresponds to the radial location of the density (pressure) maximum of the torus with $s_0=0$ for a given $a$. In each figure, a comparison with the Schwarzschild BH ($a=0$) is presented. } \label{fig-3}
  \end{figure*}

\begin{figure*}[htb!]
\captionsetup{justification=raggedright, singlelinecheck=on}
\hspace{-2cm}
\begin{subfigure}[b]{0.4\textwidth}
    \includegraphics[scale=0.5]{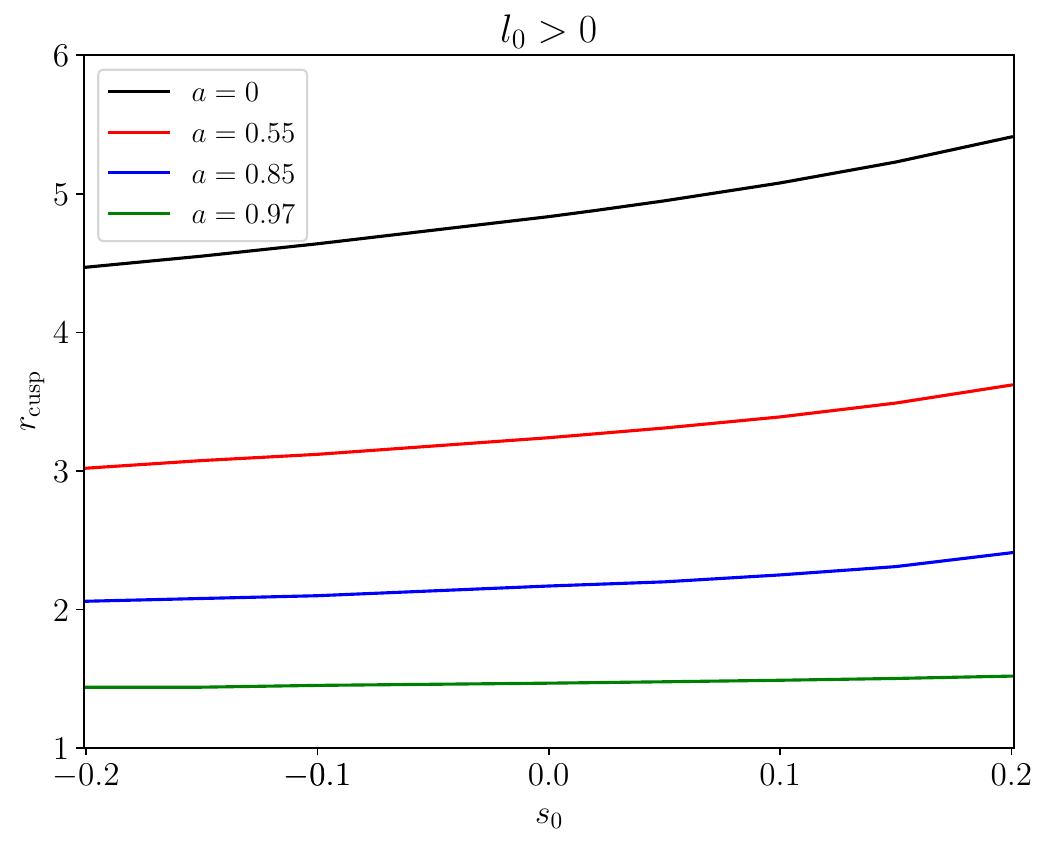}
\end{subfigure}
   \hspace{2cm}
  \begin{subfigure}[b]{0.4\textwidth}
  	\includegraphics[scale=0.5]{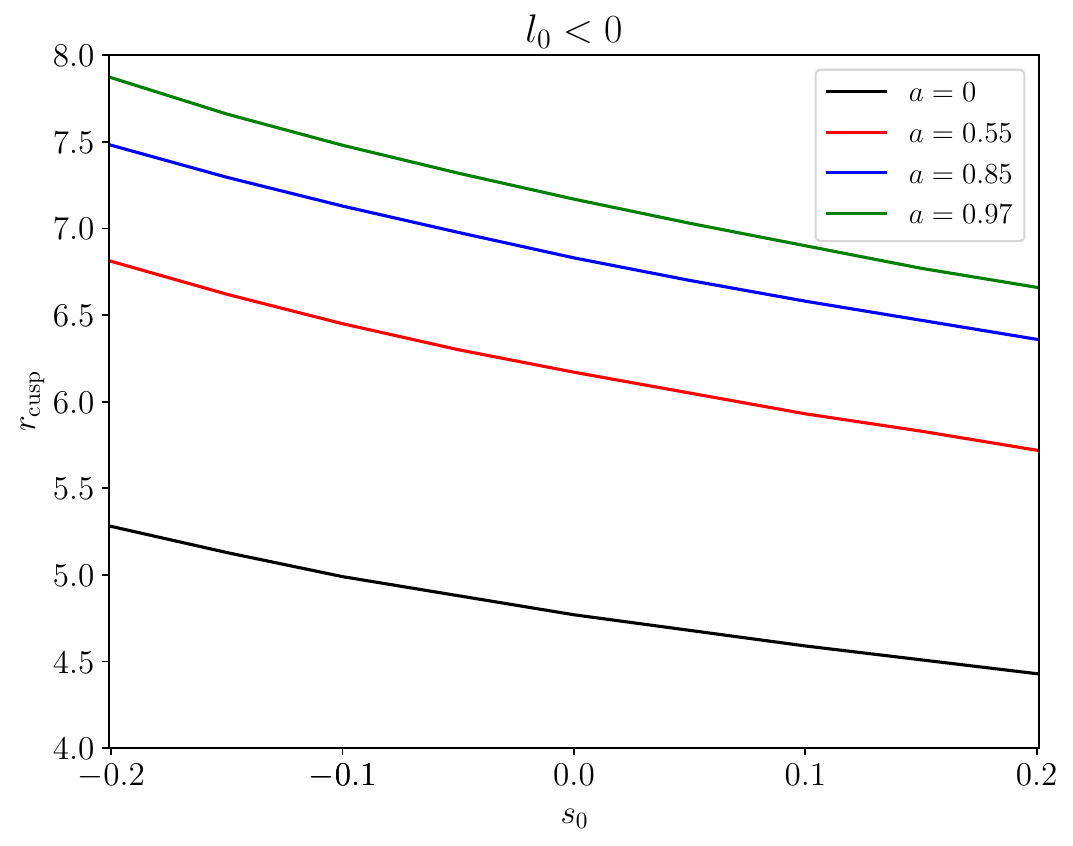}
    \end{subfigure}
    \caption{Variation of $r_{\mathrm{cusp}}$ with respect to $s_0$ for different values of Kerr spin parameter $a$. The left and right panels correspond to co-rotating and counter-rotating cases respectively. The Schwarzschild case is presented for comparison in each panel. } \label{fig-4}
  \end{figure*}

Not only $\epsilon_{\mathrm{max}}$, the radial locations of the center $r_{\mathrm{max}}$ and the cusp $r_{\mathrm{cusp}}$ move due to spin of the fluid, first noted in Figure~\ref{fig-rc} corresponding to a given value of $a$. In Figure~\ref{fig-3} and Figure~\ref{fig-4}, we respectively present the behaviour of $r_{\mathrm{max}}$ and $r_{\mathrm{cusp}}$ for a range of $s_0$ involving both positive and negative values (involving $s_0=0$) while considering a set of fixed BH spin parameter. For the co-rotating case, the left panel of Figure~\ref{fig-3} shows that $r_{\mathrm{max}}$ shifts towards the outer horizon of the Kerr BH as $s_0$ slow increases from negative to positive values of $s_0$. The pattern is exactly reverse for the counter-rotating motion presented on the right panel of Figure~\ref{fig-3}.  In fact, corresponding to the counter-rotating motion, for a given $s_0$, $r_{\mathrm{max}}$ takes in higher values as $a$ is increased (i,e. the center of the torus moves farther away from the outer horizon with increase in BH spin). 
Similarly, the positional changes of the cusp due to variation of $s_0$ and $a$ is demonstrated in Figure~\ref{fig-4} for both corotating and counter-rotating motions of the BH. 

\begin{figure*}[htb!]
\captionsetup{justification=raggedright, singlelinecheck=on}
\hspace{-3cm}
\begin{subfigure}[b]{0.3\textwidth}
  \includegraphics[scale=0.22]{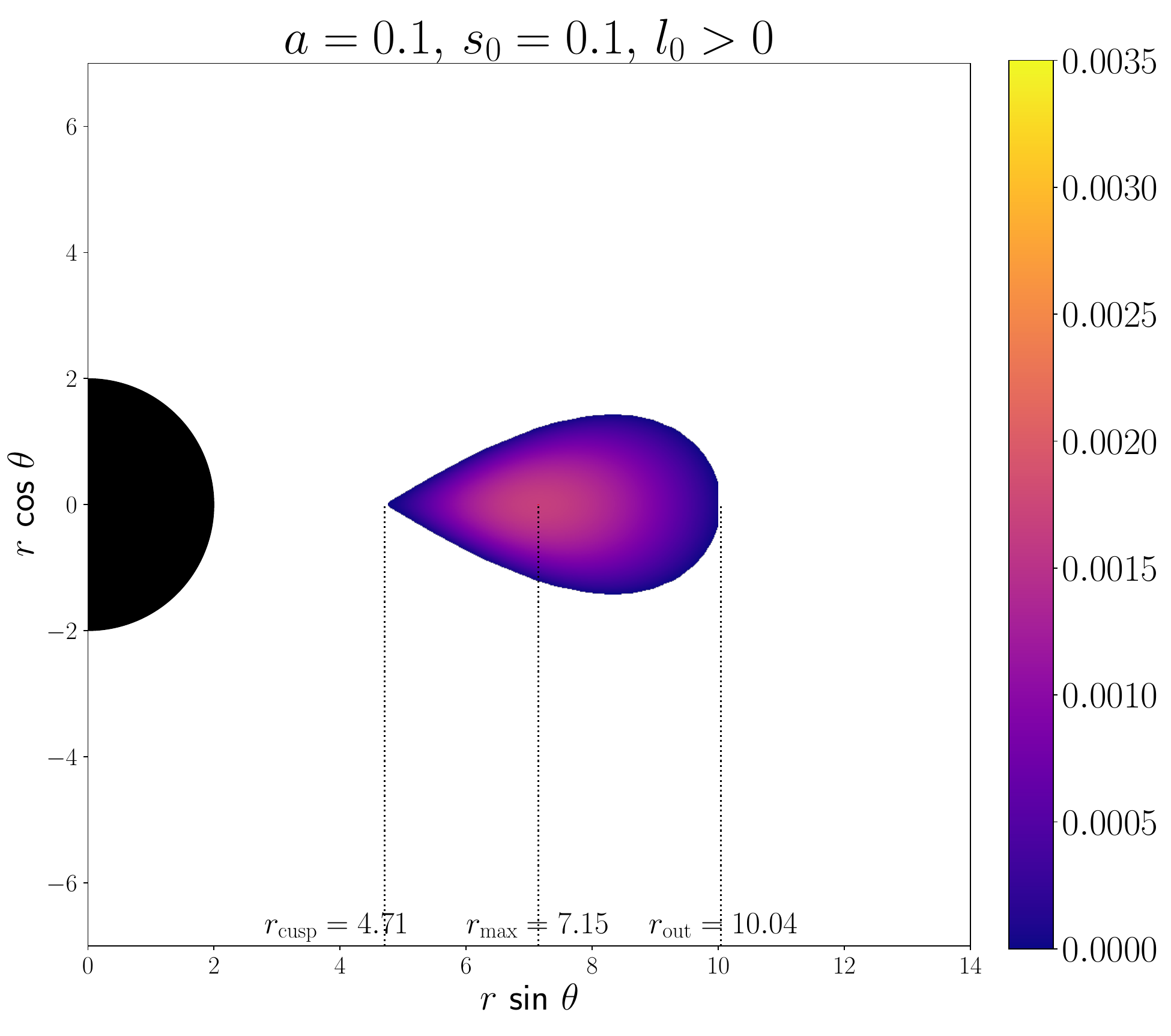}
    \end{subfigure}
    \hspace{0.6cm}
  \begin{subfigure}[b]{0.3\textwidth}
  	\includegraphics[scale=0.22]{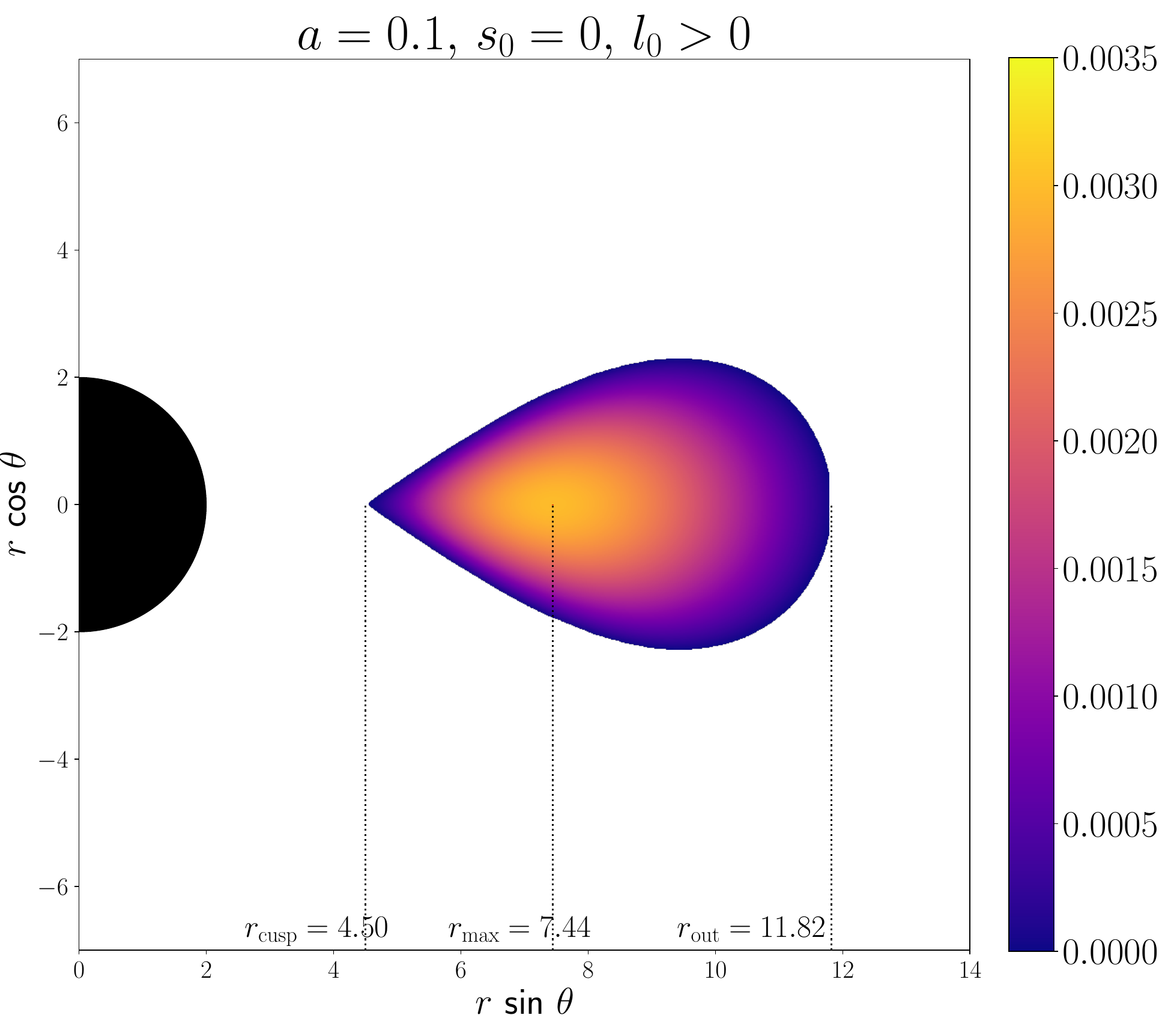}
    \end{subfigure}
    \hspace{0.6cm}
  \begin{subfigure}[b]{0.3\textwidth}
   \includegraphics[scale=0.22]{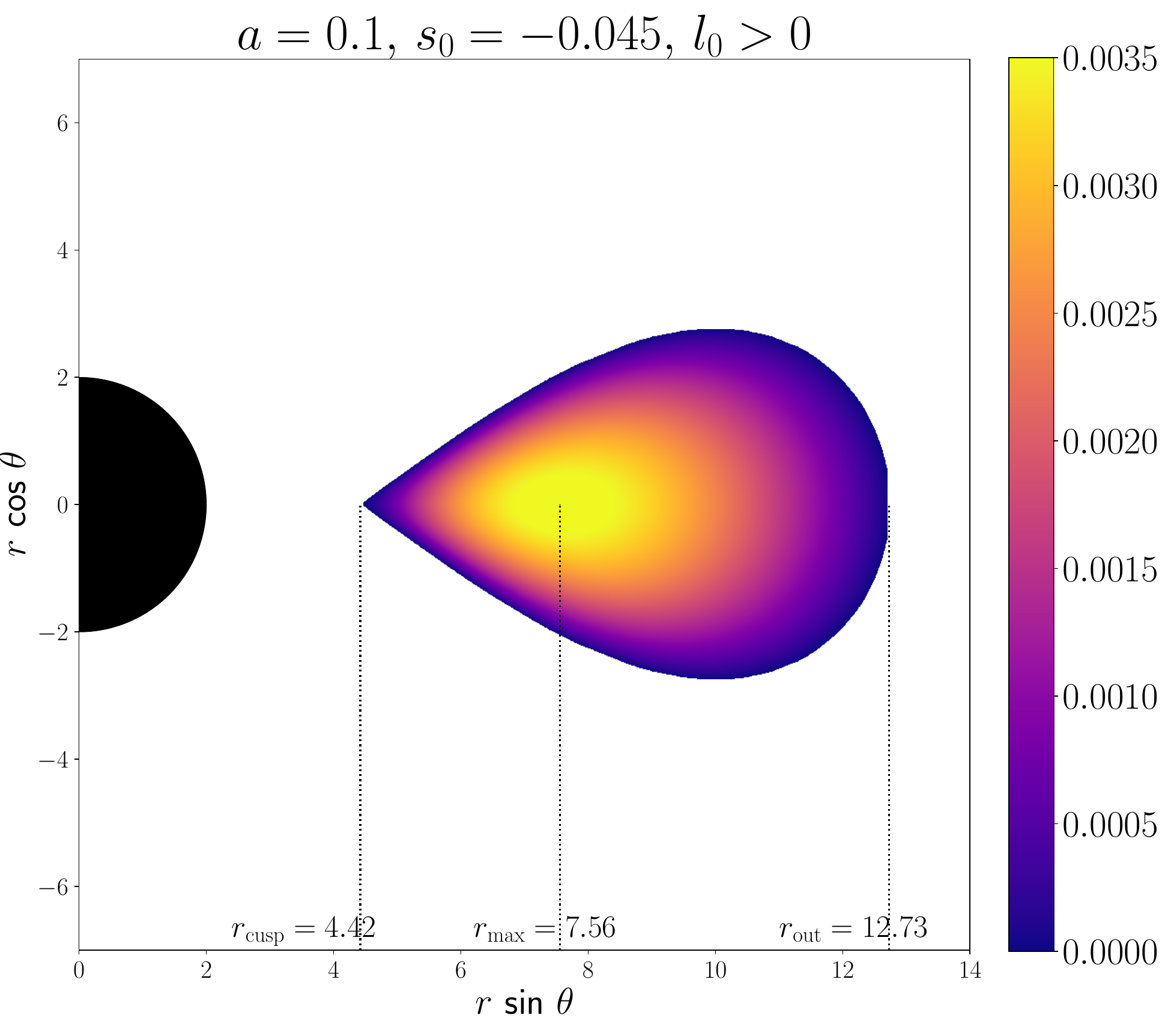}
    \end{subfigure}
\\[2mm]
    \hspace{-3cm}
    \begin{subfigure}[b]{0.3\textwidth}
   \includegraphics[scale=0.22]{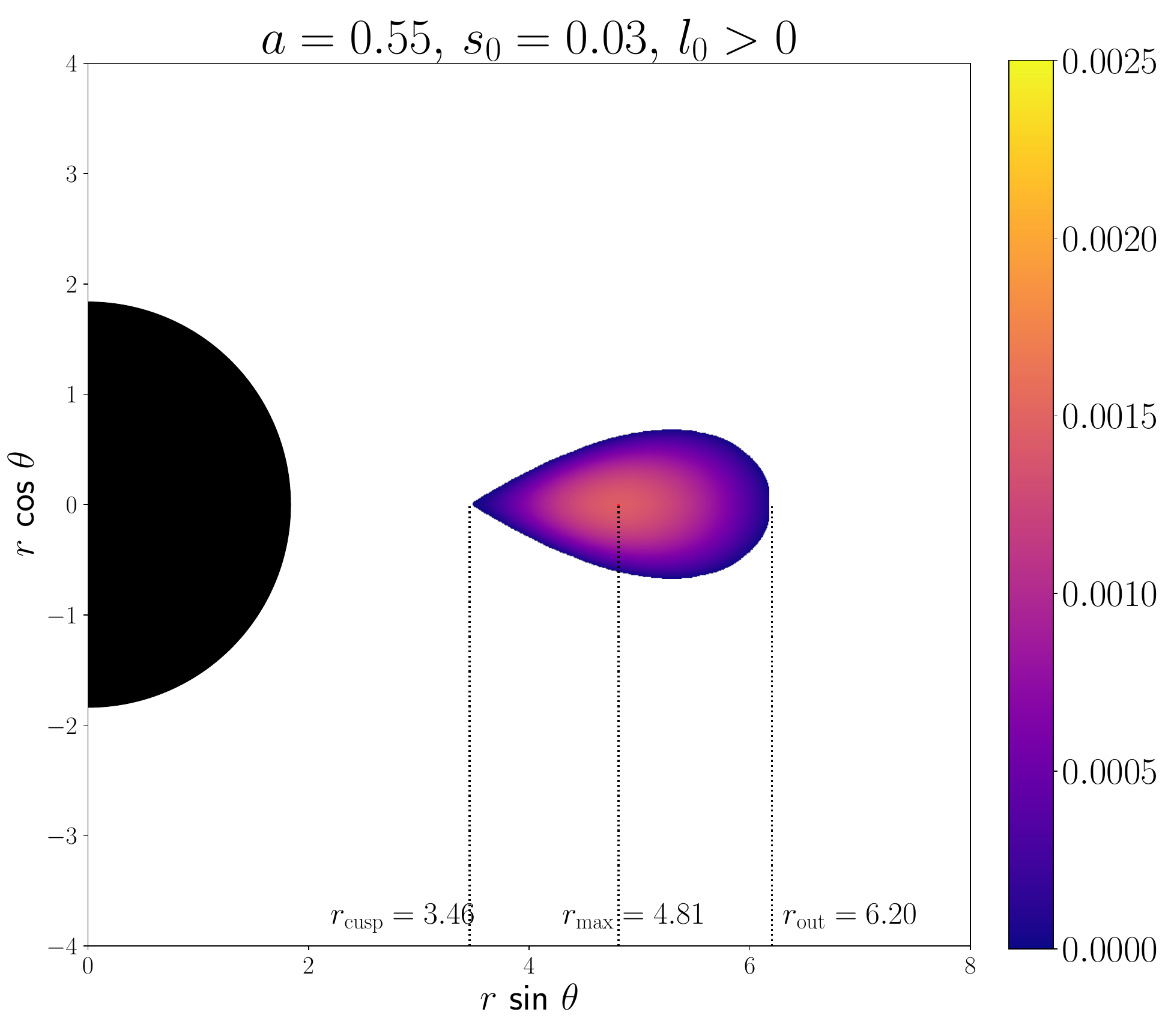}
   \end{subfigure}
   \hspace{0.6cm}
\begin{subfigure}[b]{0.3\textwidth}
\includegraphics[scale=0.22]{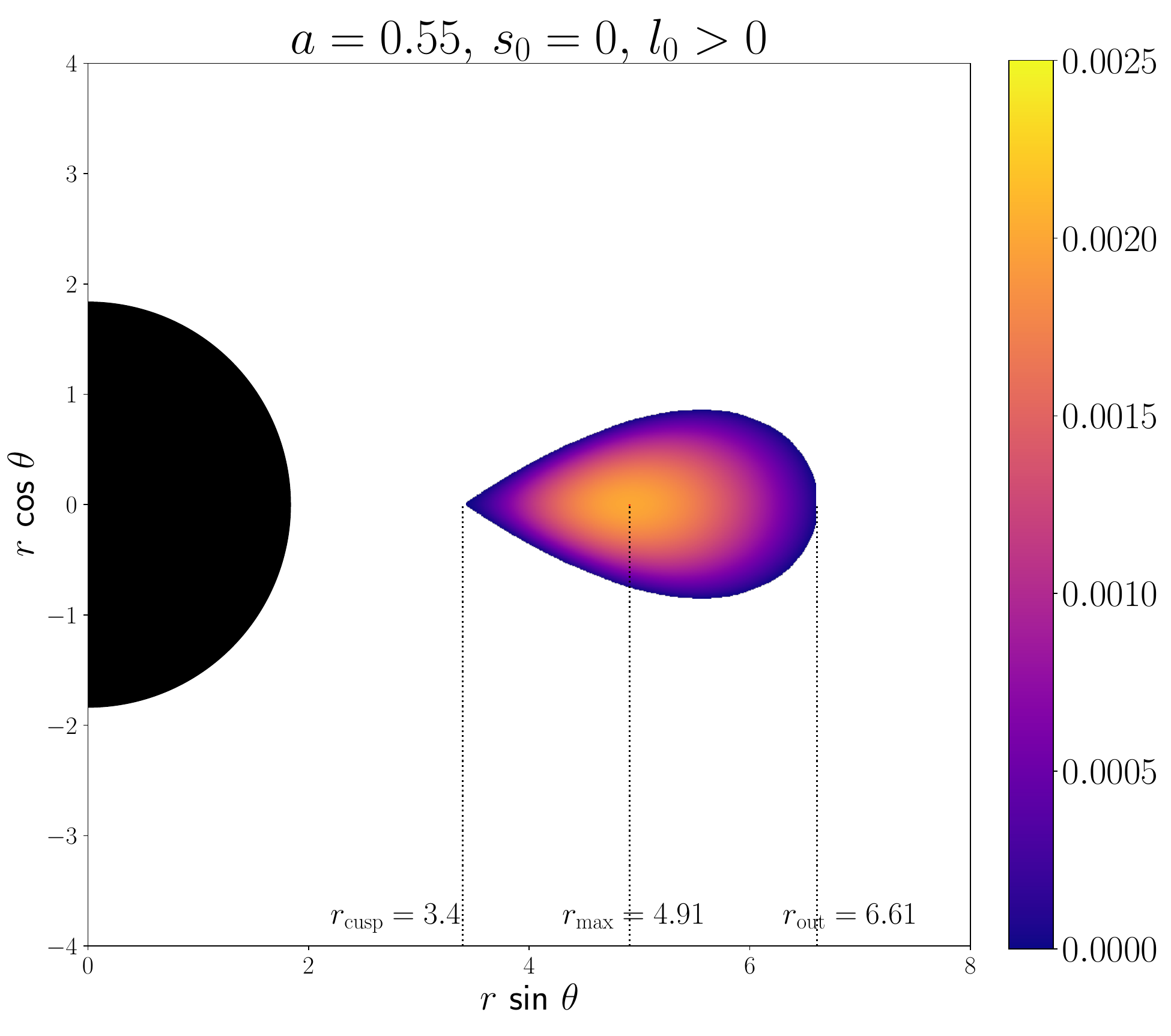}
  \end{subfigure}
    \hspace{0.6cm}
  	\begin{subfigure}[b]{0.3\textwidth}
  \includegraphics[scale=0.22]{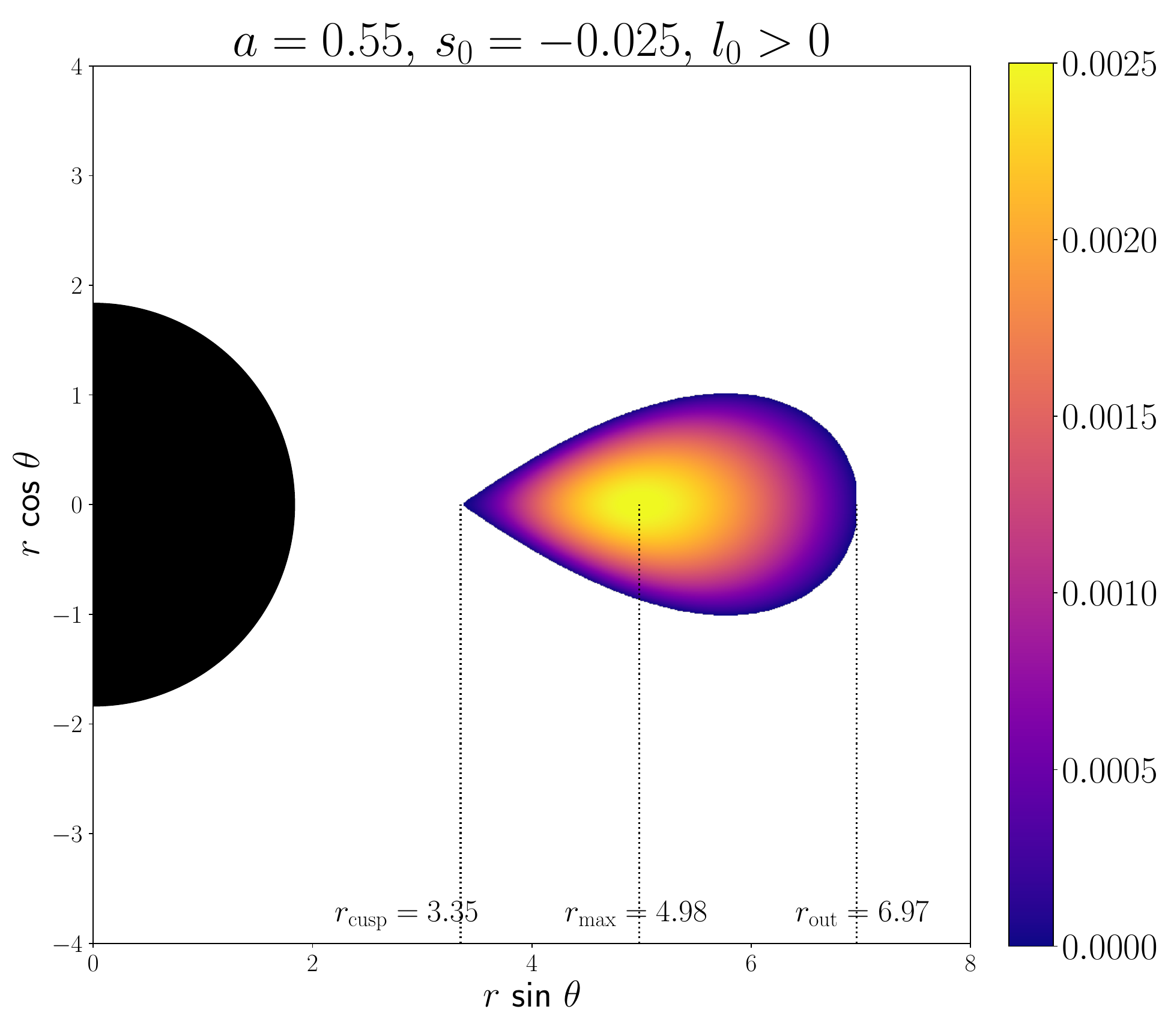}
  \end{subfigure}
  \\[2mm]
  \hspace{-3cm}
  \begin{subfigure}[b]{0.3\textwidth}
   \includegraphics[scale=0.22]{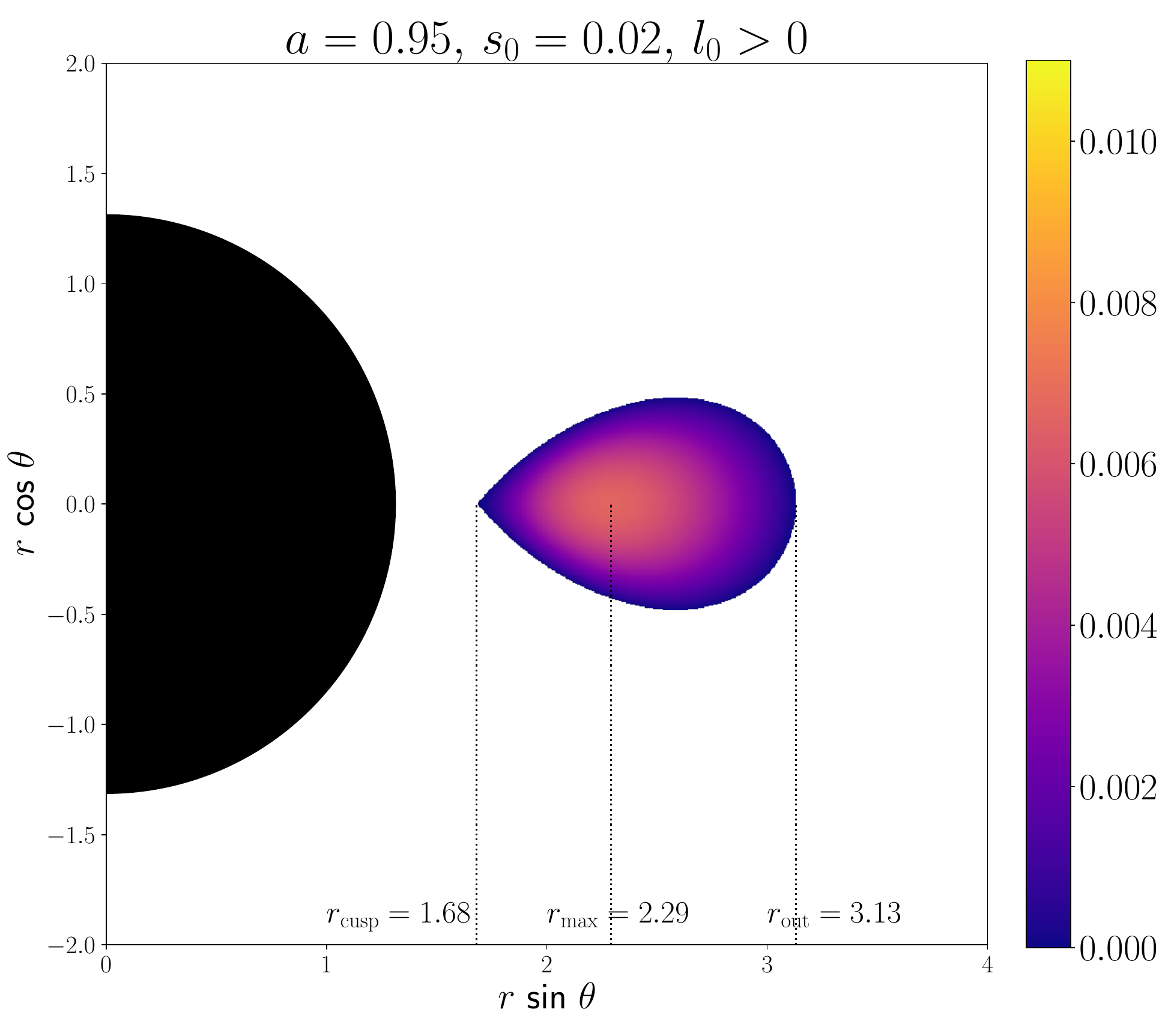}
    \end{subfigure}
   \hspace{0.6cm}
  \begin{subfigure}[b]{0.3\textwidth}
   \includegraphics[scale=0.22]{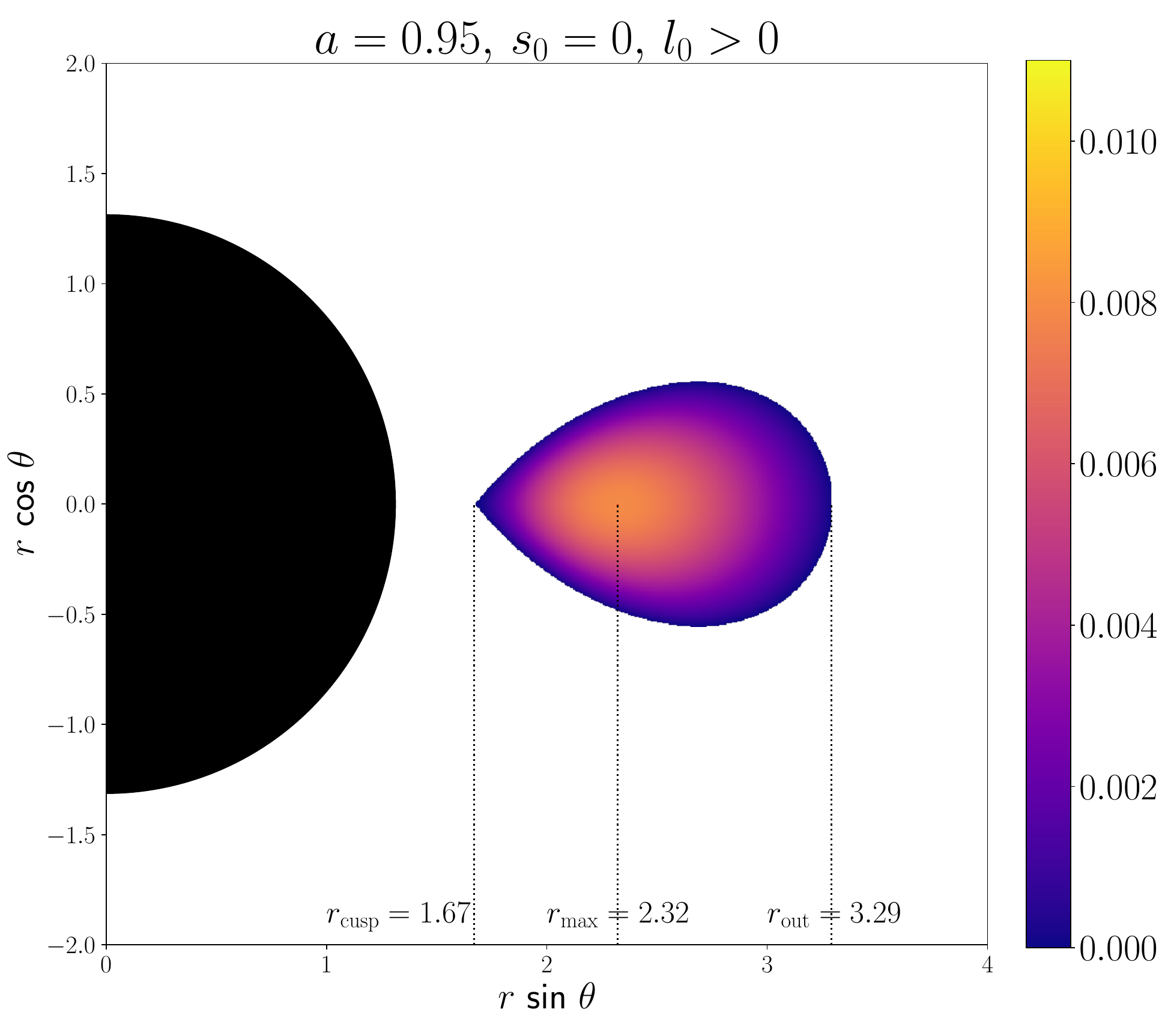}
    \end{subfigure}
    \hspace{0.6cm}
  \begin{subfigure}[b]{0.3\textwidth}
   \includegraphics[scale=0.22]{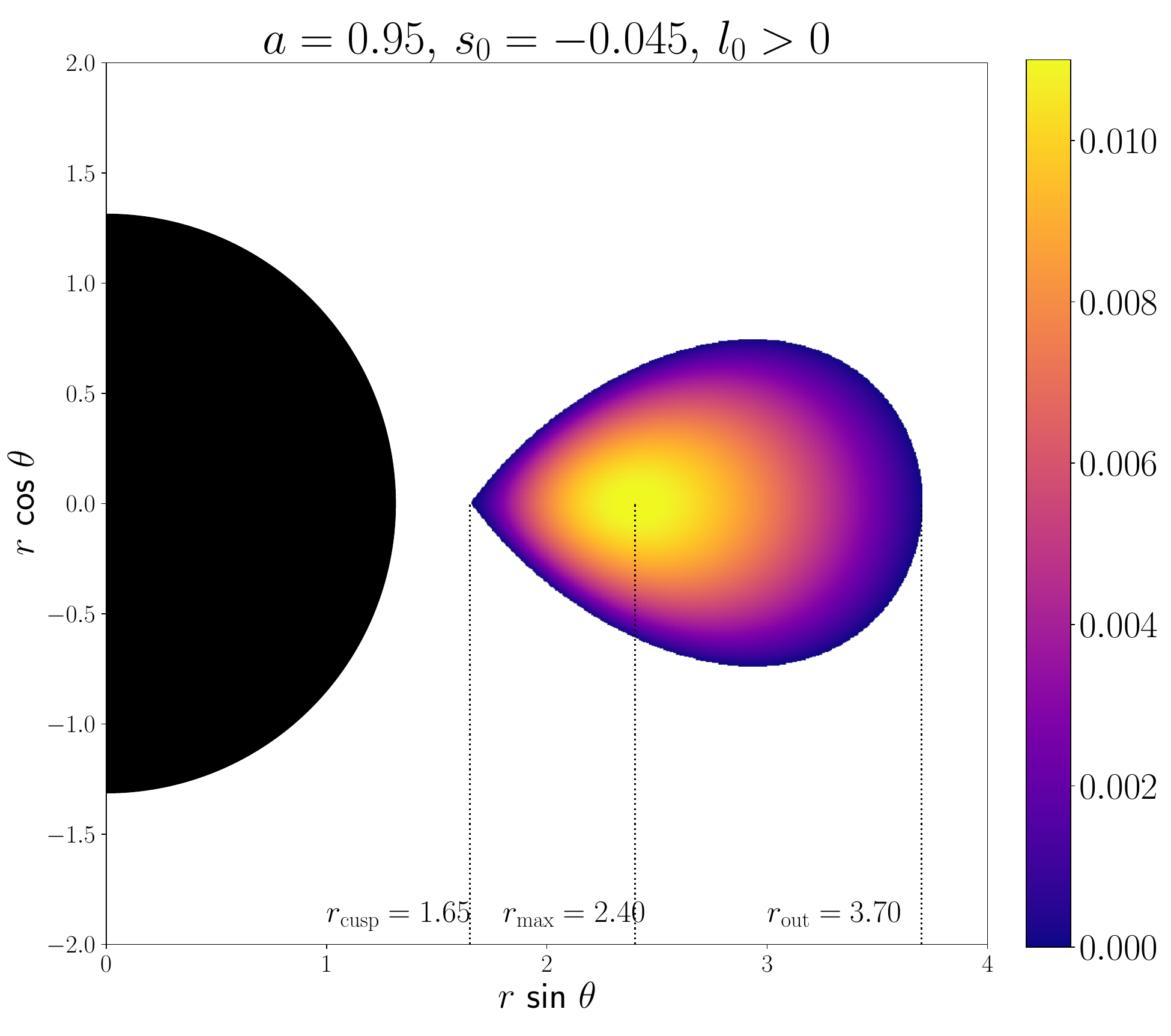}
    \end{subfigure}
    \caption{Closed equidensity surfaces of the spin fluid torus in the Kerr spacetime for different values of BH spin $a$ and spin parameter $s_0$ corresponding to the Weyssenhoff fluid.  The isodensity surfaces are obtained with increasing values of $a$ from the top to the down row and decreasing value of $s_0$ from left to right stating from positive to negative values of $s_0$ in each row. In each case, the torus without spin fluid ($s_0=0$) is presented to facilitate a comparison with torus supported by Weyssenhoff fluid. The three values of BH spin and specific angular momentum are respectively (in the increasing order of $a$) $a=0.1, l_0=3.67$, $a=0.55,l_0=3.14$ and $a=0.95, l_0=2.36$. For all the cases, we have fixed $\gamma=2$ and $M_{\mathrm{BH}}=1$.}   \label{fig-5}
   \end{figure*}

\begin{figure*}[htb!]
\captionsetup{justification=raggedright, singlelinecheck=on}
\hspace{-3cm}
\begin{subfigure}[b]{0.3\textwidth}
  \includegraphics[scale=0.22]{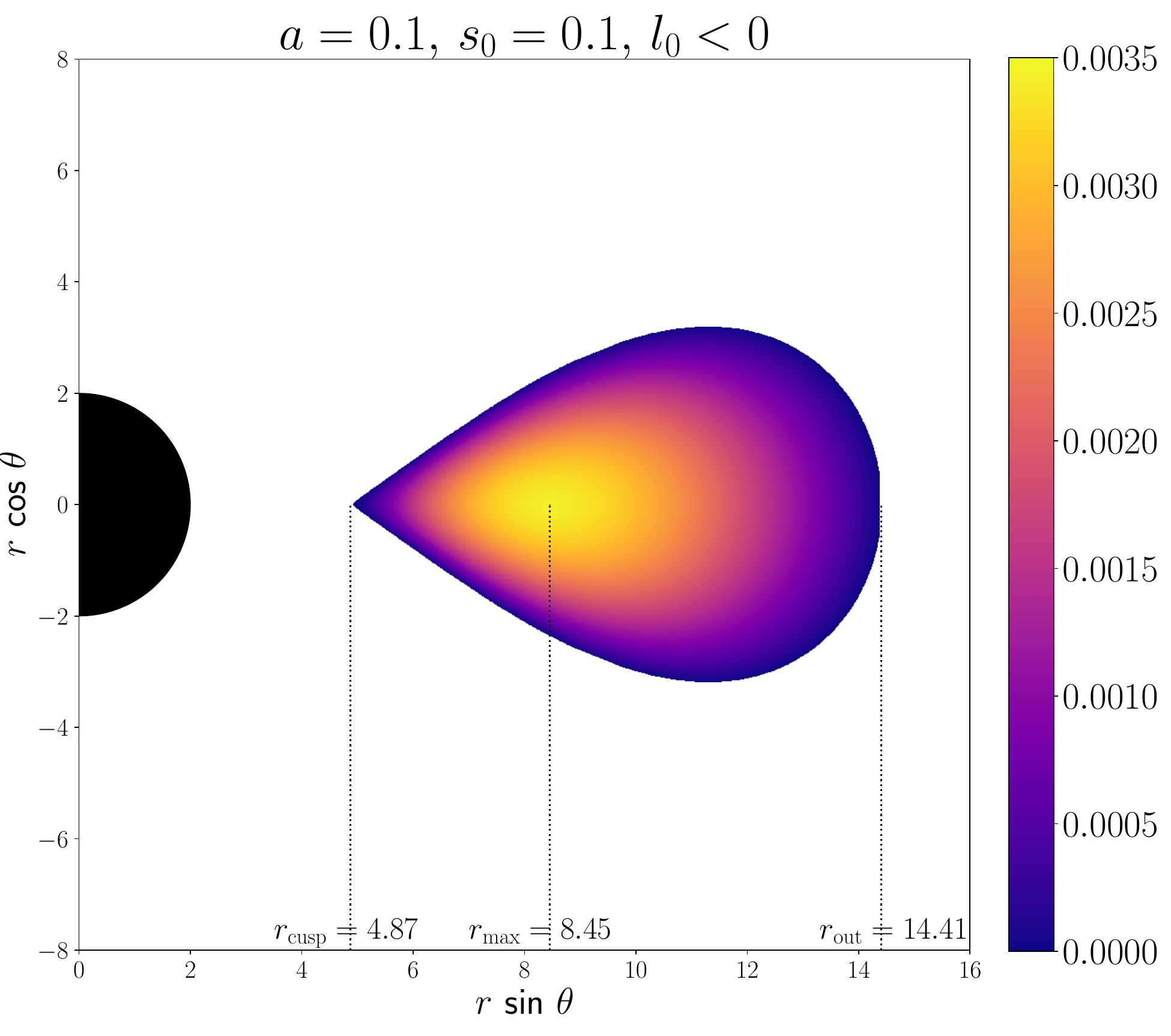}
    \end{subfigure}
    \hspace{0.6cm}
  \begin{subfigure}[b]{0.3\textwidth}
  	\includegraphics[scale=0.22]{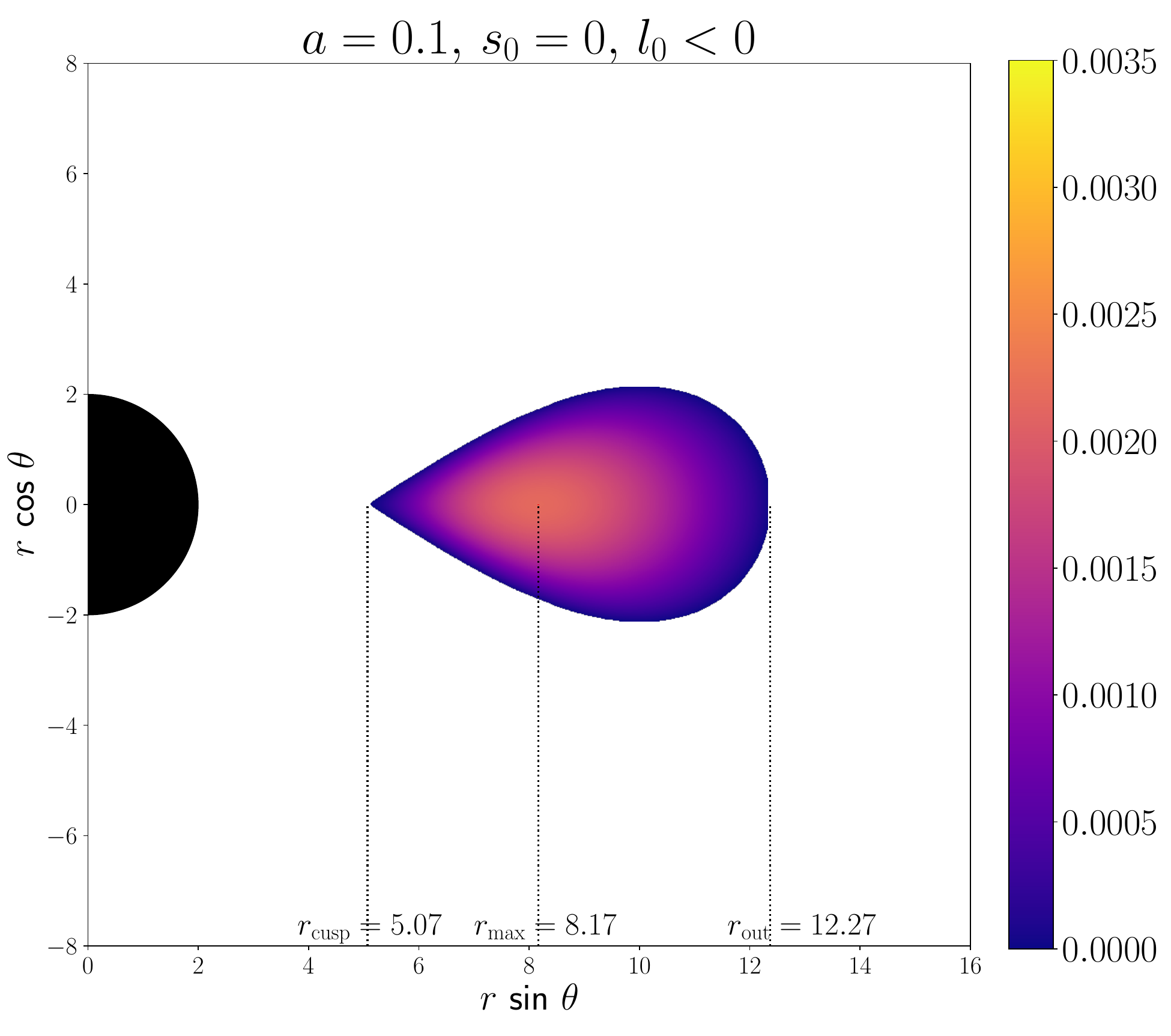}
    \end{subfigure}
    \hspace{0.6cm}
  \begin{subfigure}[b]{0.3\textwidth}
   \includegraphics[scale=0.22]{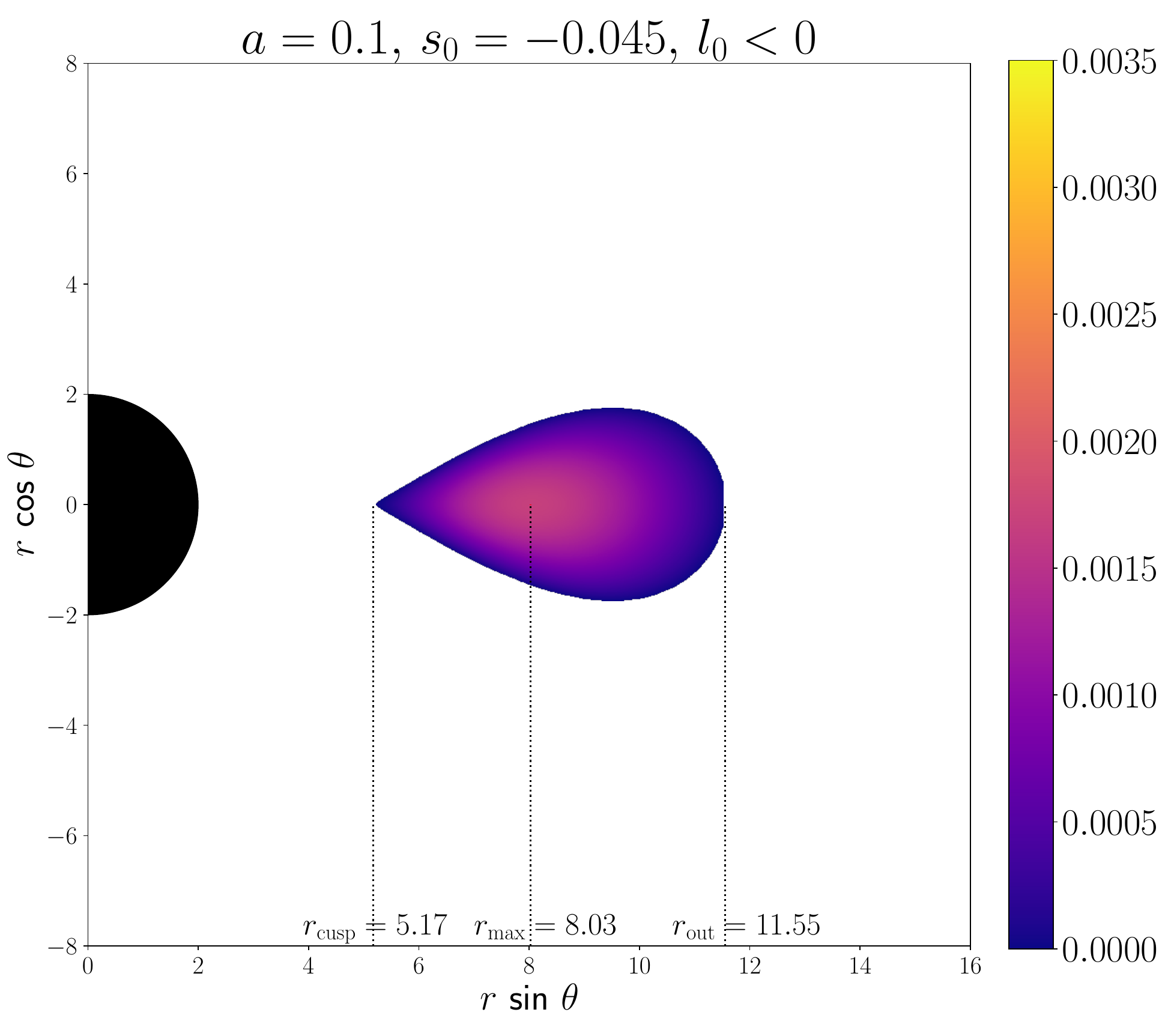}
    \end{subfigure}
\\[2mm]
    \hspace{-3cm}
    \begin{subfigure}[b]{0.3\textwidth}
   \includegraphics[scale=0.22]{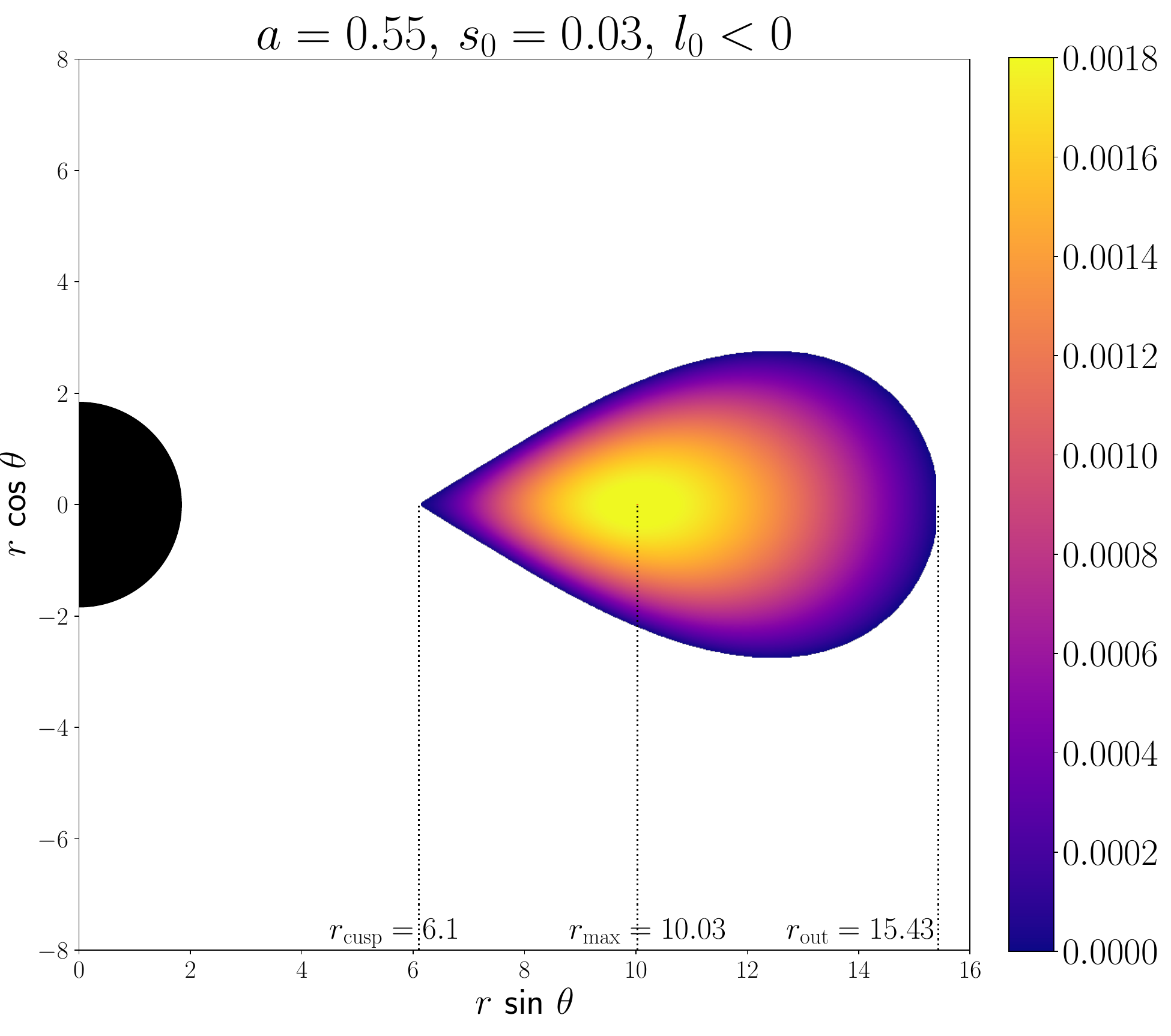}
   \end{subfigure}
   \hspace{0.6cm}
\begin{subfigure}[b]{0.3\textwidth}
\includegraphics[scale=0.22]{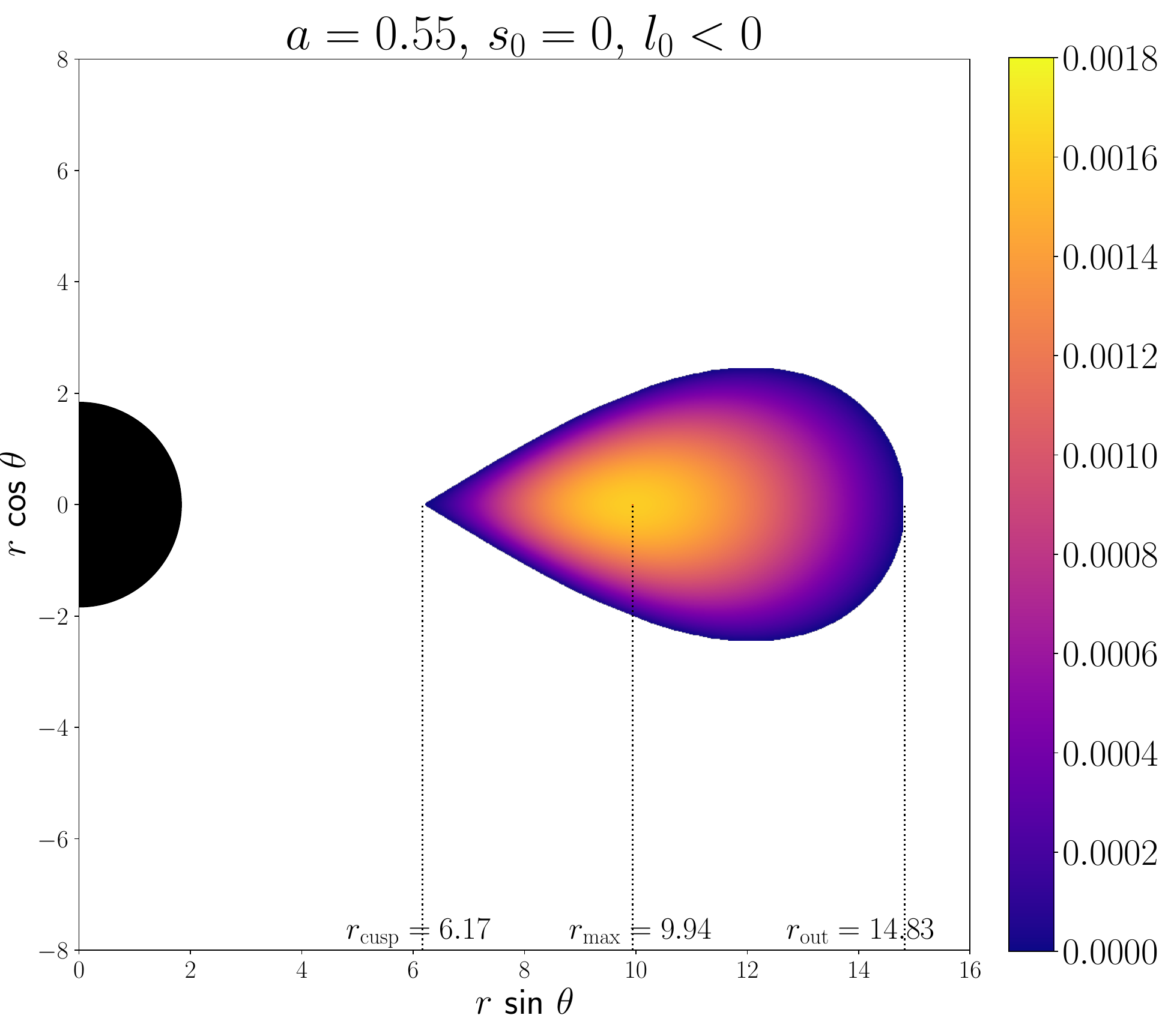}
  \end{subfigure}
    \hspace{0.6cm}
  	\begin{subfigure}[b]{0.3\textwidth}
  \includegraphics[scale=0.22]{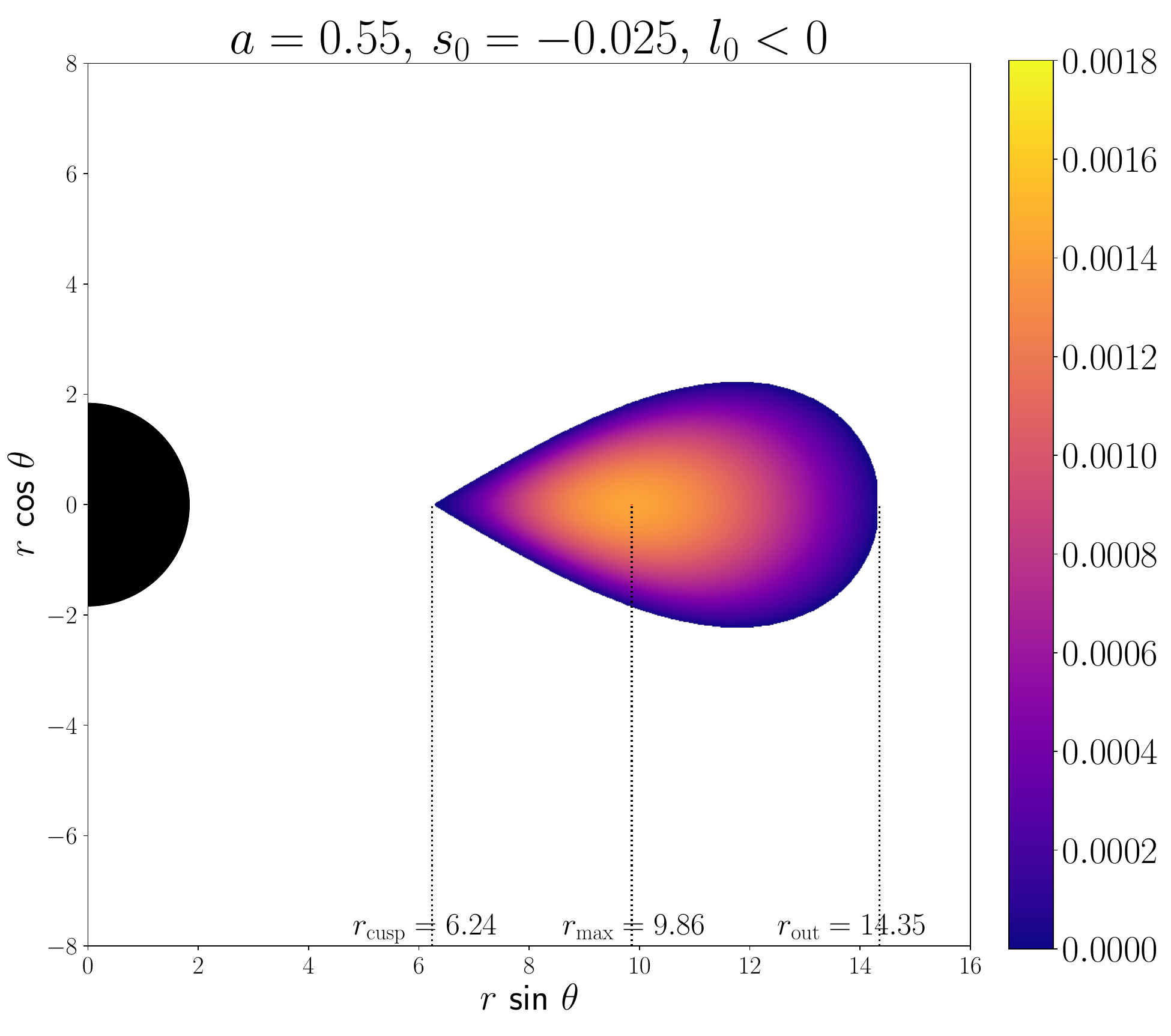}
  \end{subfigure}
  \\[2mm]
  \hspace{-3cm}
  \begin{subfigure}[b]{0.3\textwidth}
   \includegraphics[scale=0.22]{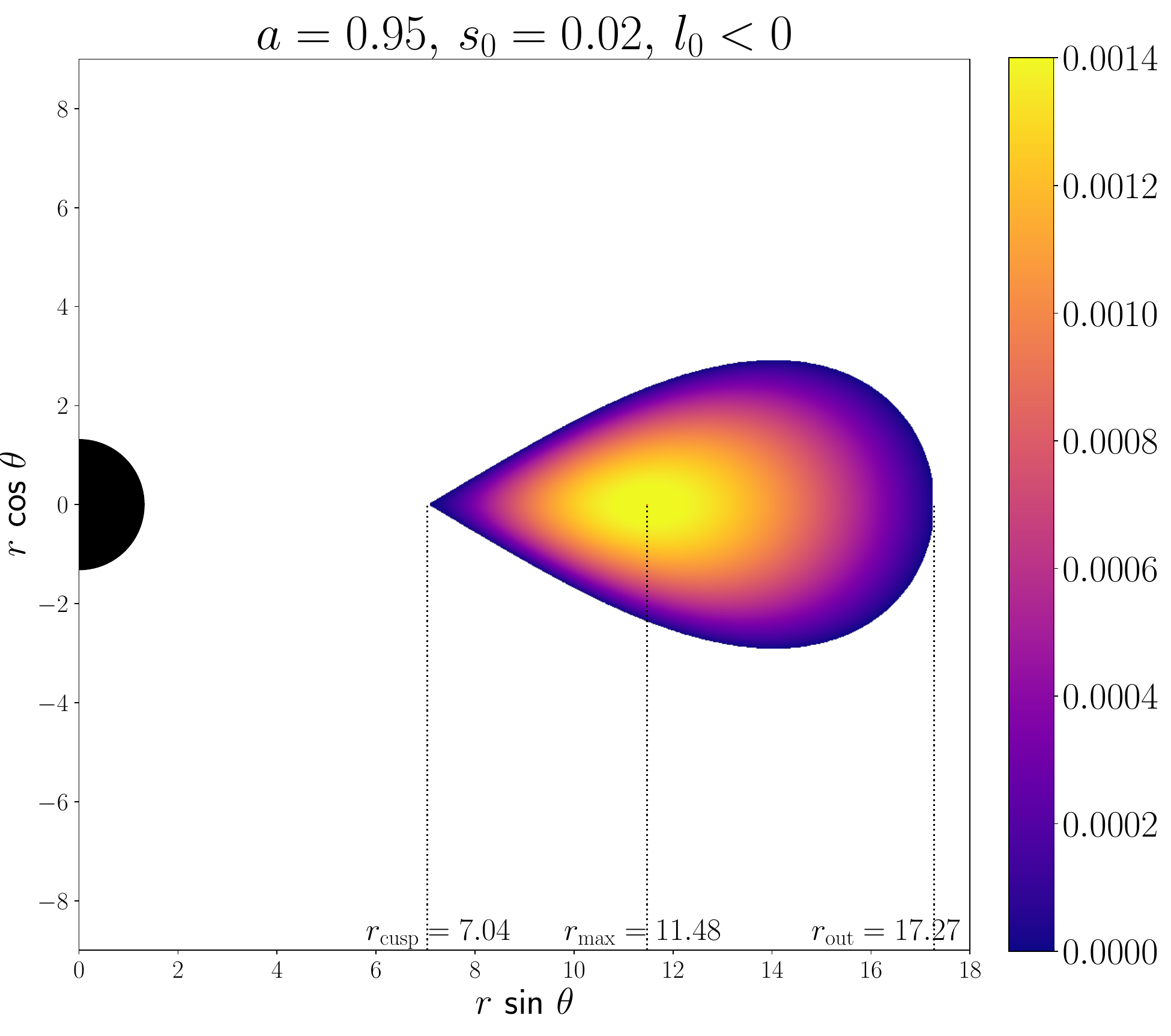}
    \end{subfigure}
   \hspace{0.6cm}
  \begin{subfigure}[b]{0.3\textwidth}
   \includegraphics[scale=0.22]{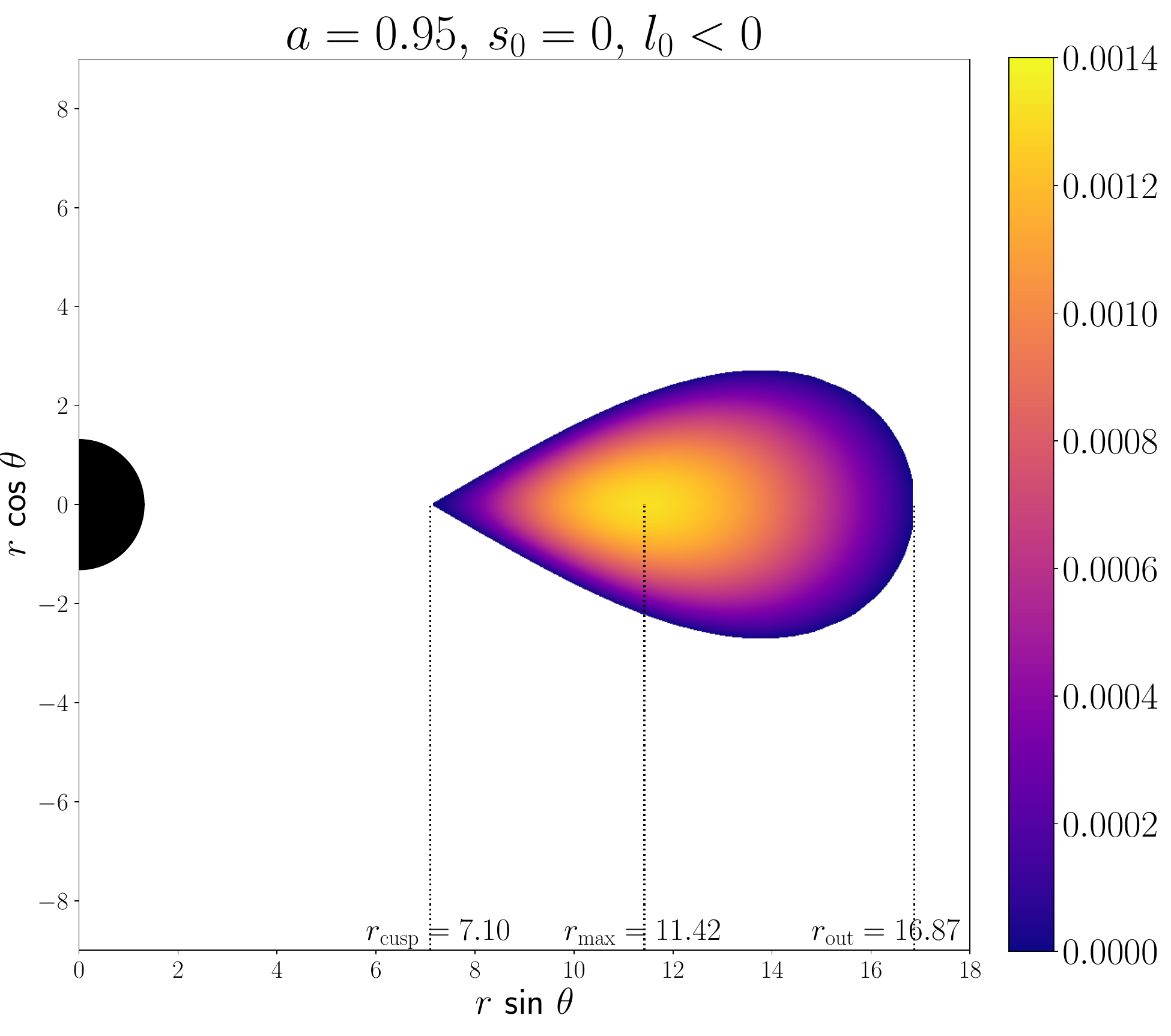}
    \end{subfigure}
    \hspace{0.6cm}
  \begin{subfigure}[b]{0.3\textwidth}
   \includegraphics[scale=0.22]{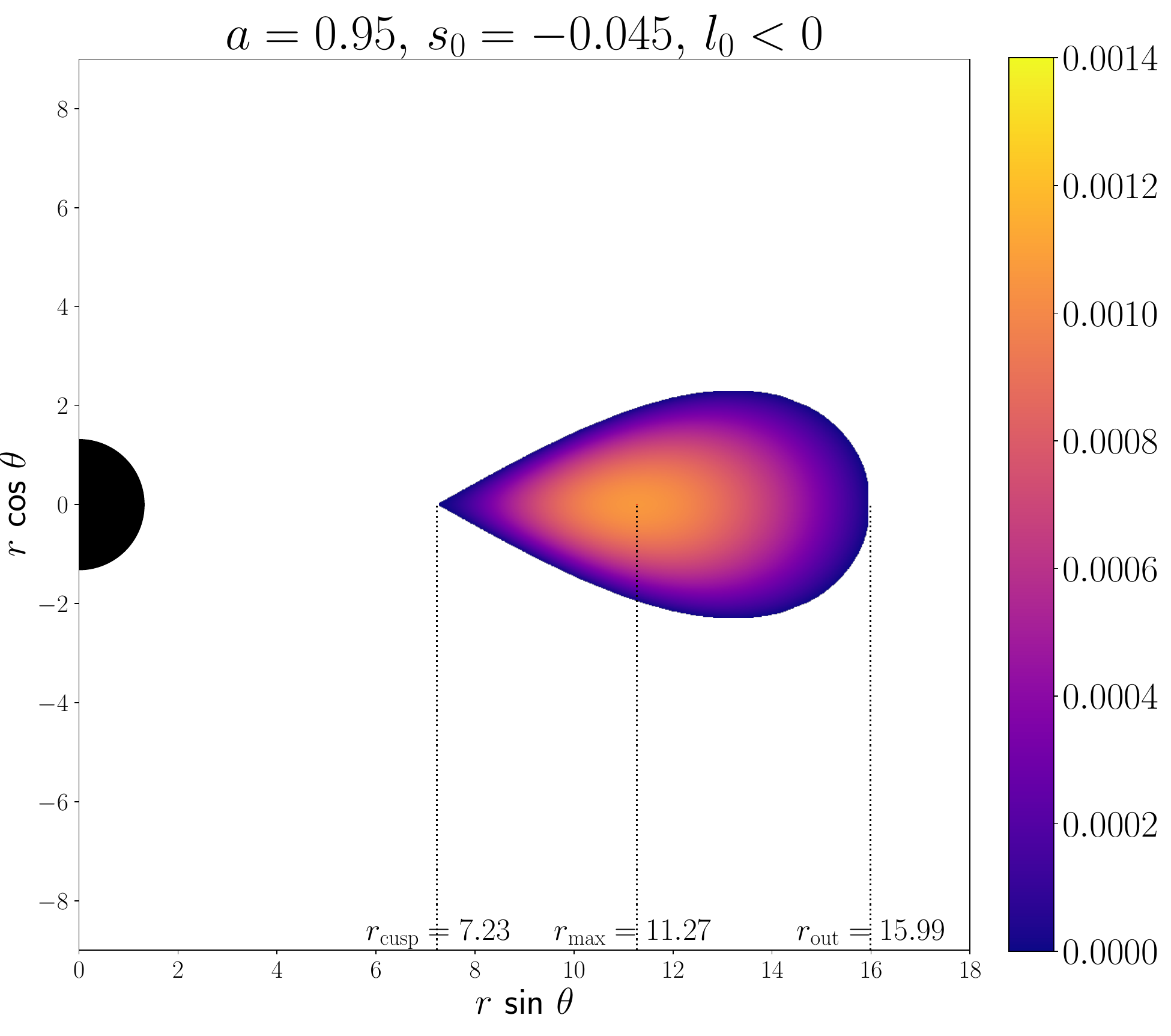}
    \end{subfigure}
    \caption{Closed equidensity surfaces of Weyssenhoff spin fluid torus in the Kerr spacetime for different values of BH spin $a$ and spin parameter $s_0$ for retrograde motion.  The three values of BH spin and specific angular momentum are respectively (in the increasing order of $a$) $a=0.1, l_0=-3.84$, $a=0.55,l_0=-4.19$ and $a=0.95, l_0=-4.46$ with fixed $\gamma=2$ and $M_{\mathrm{BH}}=1$ for all cases.}   \label{fig-6}
   \end{figure*}

Finally, the closed isodensity surfaces of the Weyssenhoff spin fluid torus are presented in Figure~\ref{fig-5} and Figure~\ref{fig-6} for corotating and counter-rotating cases respectively. 
We have considered all possible cases, involving $s_0>0, \, s_0<0$ and  $s_0=0$. For the corotating case,  Figure~\ref{fig-5} shows the radial size of the torus, defined as $\Delta r \equiv r_{\mathrm{out}} - r_{\mathrm{in}}$ increases as the magnitude of $s_0$ is decreased from positive to negative values resulting into systematic increase in the total energy density.  Notably, the feature is observed for both slowly and highly rotating cases. Nevertheless, the overall size of the torus is found to shrink significantly as one moves from a slowly to a highly  rotating spacetime, for instance, with $a=0.95$, $\Delta r $ significantly gets decreased in comparison to the  $a_0=0.1$ case. Further, with reference to the previous figures, the cusp comes closer to the event horizon of the Kerr BH whereas the center moves away with changing $s_0$ and fixed $a$.
The features of isodensity profiles of the spin fluid torus corresponding to counter-rotating motion of Kerr BH are completely opposite as presented in Figure~\ref{fig-6}. It can be readily observed that overall size of the torus as well as  $\Delta r$ diminish with decreasing $s_0$ from positive to negative values. Consequently, the energy density decreases, a feature observed for both slowly and highly rotating Kerr BH. Notably, all these characteristics of isodensity solutions were first encountered for radial solutions at the equatorial plane.

\subsection{Limits of the parameter space}
\begin{figure*}{htb!}
\centering
\includegraphics[scale=0.21]{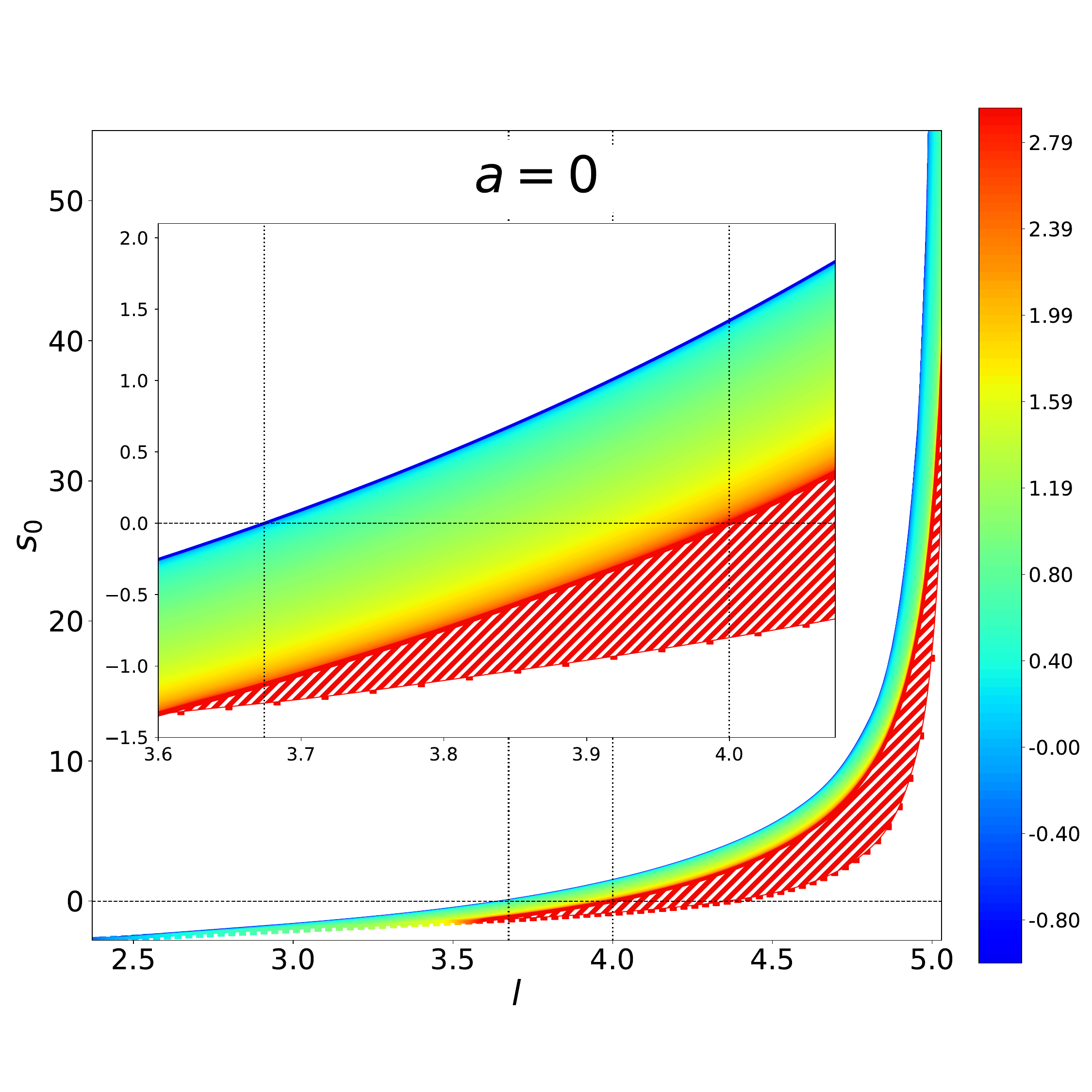}
\hspace{-0.1cm}
\includegraphics[scale=0.20]{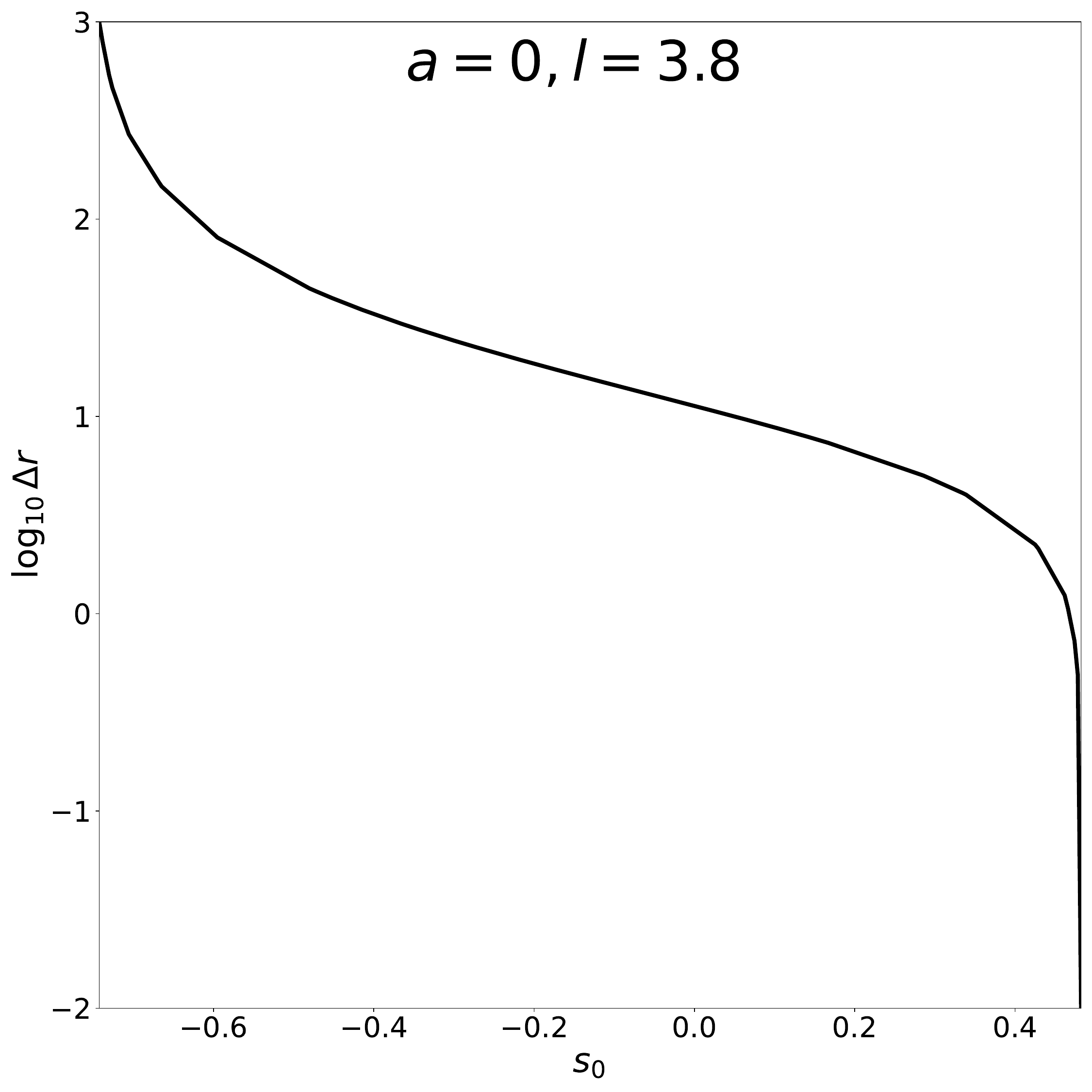}
\\
\includegraphics[scale=0.21]{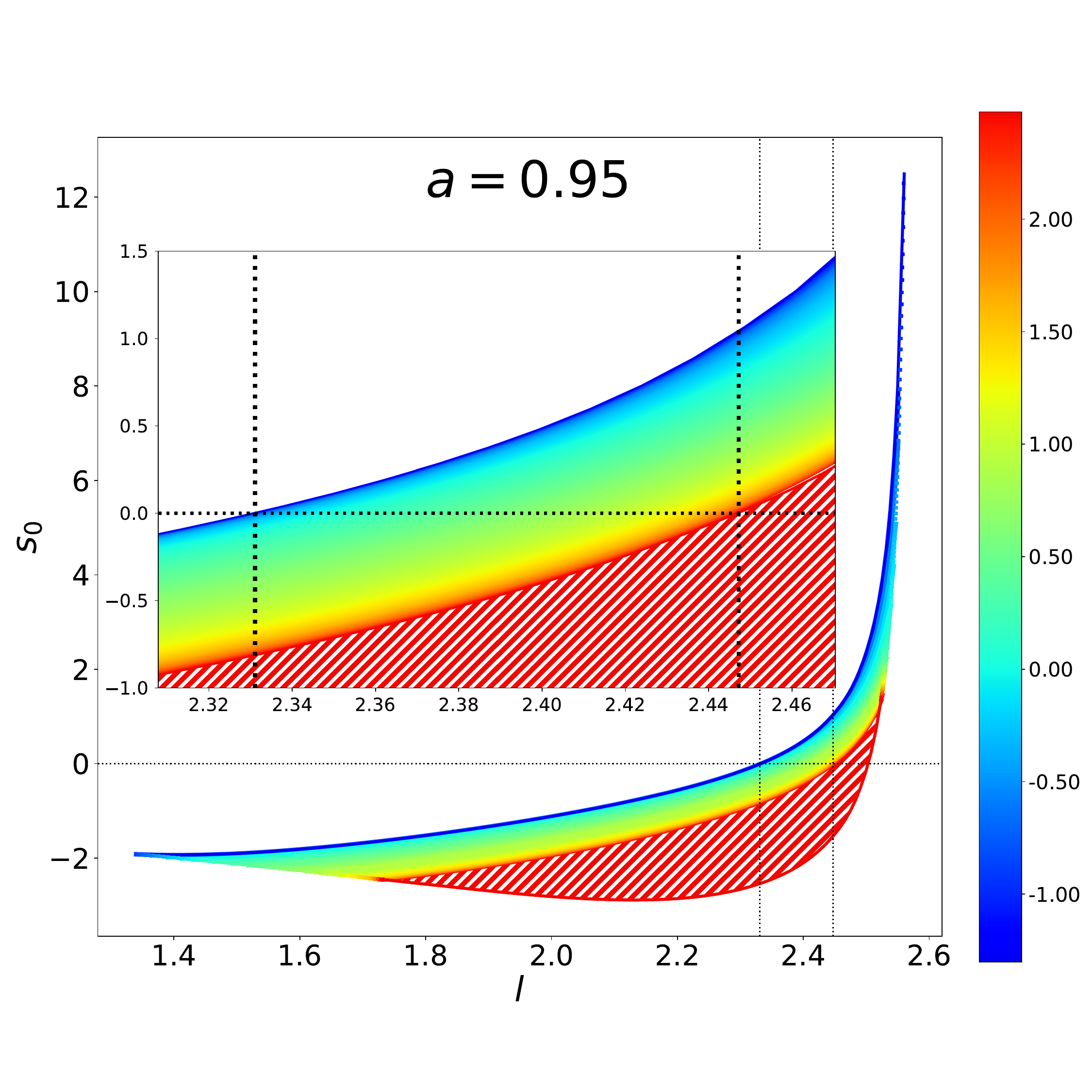}
\hspace{-0.1cm}
\includegraphics[scale=0.20]{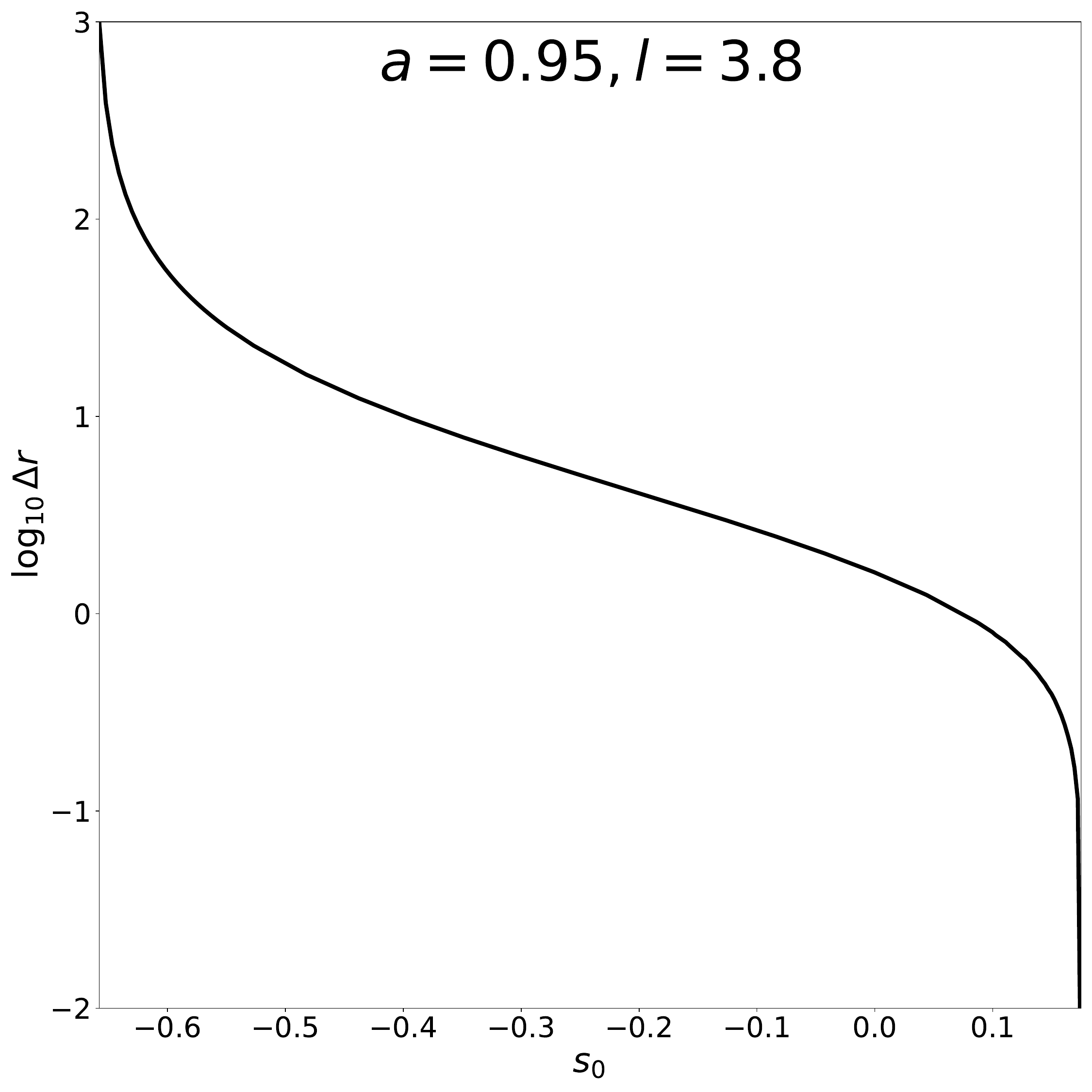}
\caption{Parameter space existence region for the Schwarzschild case $a= 0$ (top row) and for a highly rotating Kerr BH $a = 0.95$. In the Kerr case, the disk is co-rotating with the BH. In the panels at the left we have the existence region for each one of the cases. In the vertical axis we show the spin fluid parameter $s_0$, in the horizotal axis we show the specific orbital angular momentum of the disk $l_0$ and in color code, the logarithm of the radial extent of the disk $\Delta r$. In the white region there are no spin fluid solutions whithin our approach, the red, dashed regions show the region in which our code return a solution, but said solution violates one of the boundary conditions for $\epsilon$ (namely $\epsilon = 0$ at $r = r_{\mathrm{cusp}}$ or $\epsilon \rightarrow \epsilon_0 \neq 0$ for $r \rightarrow \infty$). The vertical black dotted lines represent the values of $l_{\mathrm{ms}}$ and $l_{\mathrm{mb}}$ for a spin-less fluid and the horizontal black, dotted line represent $s_0 = 0$. In the inset, we focus in the region of the parameter space where the specific angular momentum $l_0$ is close to the interval $l_{\mathrm{ms}} < l_0 < l_{\mathrm{mb}}$.In the panels at the right, we have a vertical cut of the plots at the left, to highlight the dependence of the radial extent of the the disk $\Delta r$ (in logarithmic scale) with the parameter $s_0$.}
\label{parameters_1}
\end{figure*}

As we shown earlier, we can see that the presence of spin modifies the size of the disk and the location of its characteristic radii, so it is to be expected that, if the value of the parameter $s_0$ goes too high or too low for some given values $(a, l_0)$, we will encounter a point in which the disk becomes too big, or ceases to exist. Here we are going into more detail regarding that topic. Firstly, we want to remark that we are going to restrict the discussion to the distribution of energy density $\epsilon$ at the equatorial plane. 
Furthermore, we will also restrict this discussion to co-rotating disks.\\
 As it is known from the standard theory of thick accretion disks, a disk with specific angular momentum equal to the Keplerian angular momentum at the marginally bound orbit $l_0 = l_{\mathrm{mb}}$ is marginally infinite in radial extension (the radial extension of the disk as defined by the quantity $\Delta r \equiv r_{\mathrm{out}} - r_{\mathrm{in}}$), this means that the isobaric surface that crosses the cusp is closing exactly at infinity (and that a surfaces that crosses the equatorial plane at a radius $r > r_{\mathrm{cusp}}$ will close at a finite radius). This fact tells us that the $s_0 = 0$ solution is already marginally infinite, so we cannot use negative values of $s_0$ without violating the boundary conditions for $\epsilon$ and therefore, we only admit positive values of $s_0$. Conversely, a disk with specific angular momentum equal to the Keplerian angular momentum at the marginally stable orbit $l_0 = l_{\mathrm{ms}}$ has zero extension and therefore, the $s_0 = 0$ solution collapses to a point. To have disks of finite size we must restrict ourselves to negative values of $s_0$. A singular situation that can appear in the spin fluid scenario is that we have finite sized disks for values of $l_0 > l_{\mathrm{mb}}$ and also for values of $l_0 < l_{\mathrm{ms}}$, something that cannot happen in the standard theory. However, one could ask if it is possible to vary $s_0$ and $l_0$ in a way that one could obtain valid solutions for any value of $l_0$ and the answer seems to be no. In Fig.~\ref{parameters_1} we show the parameter space existence diagram for two different values of the Kerr spin parameter, in the top row for a Schwarzschild BH ($a = 0$) and in the bottom row for a highly rotating Kerr BH ($a = 0.95$). The color code shows, in logarithmic scale, the radial extent of the disk for a certain set of parameters $(l_0, s_0)$. The bright red color represent infinite disks (in the code, disks with $\Delta r >10^3$). The deep blue color represent infinitely small disks  (in the code, disks with $\Delta r < 10^{-2}$). The red, dashed regions show the region in which our code return a solution, but said solution violates one of the boundary conditions for $\epsilon$, namely $\epsilon = 0$ at $r = r_{\mathrm{cusp}}$ or $\epsilon \rightarrow \epsilon_0 \neq 0$ for $r \rightarrow \infty$ one of which is an assumption of our model and the other one being a standard boundary condition for accretion disks (a fluid element at infinity is unbounded). The vertical black dotted lines represent the values of $l_{\mathrm{ms}}$ and $l_{\mathrm{mb}}$ for a spin-less fluid and the horizontal black, dotted line represent $s_0 = 0$. The white region corresponds to the no existing solution region for our approach.\\
 One thing that is easy to notice is that there are two critical values of the specific orbital angular momentum of the disk $l_0$ (one maximum and one minimum) in which one can no longer increase or decrease $l_0$ irrespective of the value of $s_0$. This happens for the two values of $a$ that we are considering here. Another noteworthy fact is that, in the low $l_0$ region ($l_0 < l_{\mathrm{ms}}$), there is a point for which, if $l_0$ is further decreased, marginally infinite disks cease to exist (irrespective of $s_0$) and the available range of $\Delta r$ becomes smaller as $l_0$ further decreases until it becomes $0$ at the critical point. In the other side of the parameter space ($l_0 > l_{\mathrm{mb}}$), what happens depends on the value of the Kerr spin parameter $a$. For the Schwarzschild case, as $l_0$ approaches the critical value, it seems that $\Delta r$ becomes independent of $s_0$, and it is possible to obtain solutions for extremely large values of $s_0$ without almost any change in $\Delta r$. However, is surprising that in the Kerr case, we observe a behavior for high values of $l_0$ very similar to the one we describes for small values of $l_0$, in which we have a maximum value of $l_0$ that allows for the existence of marginally infinite disks and the possible range of $\Delta r$ decreases until it becomes $0$.

\section{Conclusions and outlook}
In this paper, we have presented stationary equilibrium solutions of closed Weyssenhoff spin fluid torus in the background of a Kerr BH. 
By incorporating spin-curvature coupling effects, we systematically analyzed how the intrinsic spin of the fluid alters the morphology of the torus for both corotating and counter-rotating motions. To obtain the equilibrium solutions of the energy density and hence the fluid pressure, we focus to the simple scenario, in which the torus, characterised by constant orbital angular momentum distributions, is non-self gravitating and undergoes circular motion in the rotating background. Furthermore, the solutions were determined under the key considerations, (a) the macroscopic spin vector is always aligned perpendicular to the equatorial plane of the Kerr BH, (b) the Frenkel SSC is imposed, leading to two independent components of the spin tensor $S^{\mu \nu}$.  Taking barotropic equation of state with $\gamma=2$ and with the help of integrability conditions, the stationary equilibrium solutions of energy density for both corotating and counter-rotating motions of the Kerr BH are then determined numerically by solving the general relativistic momentum conservation equation. The adopted numerical technique involves the method of characteristics to solve a set of PDE for given values of Kerr spin parameter $a$, $l_0$ and fluid spin parameter $s_0$. This approach finally leads to the computation of the isodensity surfaces of spin fluid torus and spin length function presented in this work for both corotating and counter-rotating cases. 

The isodensity solutions clearly demonstrate the cumulative effects of Kerr spin $a$ and spin parameter of the fluid, $s_0$ on the morphology of the torus. To summarise, for the corotating cases negative values of $s_0$ enlarge the torus and increase its peak energy density, while positive values produce more compact configurations with lower energy densities. Conversely, the counter-rotating case reveals an opposite trend: positive $s_0$ leads to denser and larger tori, whereas negative $s_0$ shrinks and rarefies them. These trends persist across both moderately and highly spinning Kerr spacetimes and are consistently reflected in radial energy density profiles, spin length function, and isodensity surfaces.
Additionally, the radial locations of the center $r_{\mathrm{max}}$ and cusp $r_{\mathrm{cusp}}$ shift systematically with $s_0$, with the direction of movement depending on the sign of $s_0$ and the direction of orbital motion. These positional shifts become more pronounced with increasing magnitude of $a$, especially in counter-rotating cases. 
The isodensity surfaces corroborate these findings, revealing systematic thickening or contraction of the torus with changes in $s_0$, depending on the rotation direction. Notably, these effects are qualitatively similar to those found in Schwarzschild spacetime, but are quantitatively more pronounced in the Kerr case. In addition, we have also presented an analysis on allowed parameter space comprising of $a, l_0, s_0$ for the existence of density solutions for both Kerr and Schwarzschild spacetimes. For a given Kerr background and specific angular momentum, there exist bounds on $s_0$ beyond which no equilibrium solution are found, either due to the absence of a cusp or the disappearance of a pressure maximum.

Overall, our study reveals that the intrinsic spin of the disk fluid introduces notable modifications to the torus morphology. These results suggest that spin effects should be incorporated in more realistic models of compact-object accretion, particularly in the strong-field regime of rapidly rotating black holes. Future work may extend to include self-gravitating effects, dynamical evolution, and observational implications such as study of relativistic jet mechanisms in the presence of accretion flows with spin fluid. On the other hand, a more realistic scenario would be to incorporate non-constant angular momentum distributions. We leave these issues for future endeavours.

\begin{acknowledgments}
SG-S is supported by CIDMA  under the FCT Multi-Annual Financing Program for R\&D Units and 
by the projects PTDC/FIS-AST/3041/2020, CERN/FIS-PAR/0024/2021 and 2022.04560.PTDC.
Further support is provided by the EU’s Horizon 2020 
research and innovation (RISE) program H2020-MSCA-RISE-2017 (FunFiCO- 777740) and 
by the European Horizon Europe staff exchange (SE) program HORIZON-MSCA-2021-SE-01 (NewFunFiCO-10108625). 
The work of SL is partially supported by Deutsche Forschungsgemeinschaft (DFG) Grant No. 40401154. 
\end{acknowledgments}

\clearpage

\bibliographystyle{apsrev4-1}
\bibliography{accretion3}

\end{document}